\documentclass{pasj00}

\begin{document}
\SetRunningHead{Y. Takeda et al.}{Sodium Abundances of A-Type Stars
from Na~I 5890/5896 Lines}
\Received{2009/06/01}
\Accepted{2009/07/01}

\title{Can Sodium Abundances of A-Type Stars Be Reliably Determined \\
from Na~I 5890/5896 Lines?
\thanks{The electronic table (table E) will be made available 
at the PASJ web site upon publication, while it is provisionally placed at 
$\langle$http://optik2.mtk.nao.ac.jp/\~{ }takeda/Asodium/$\rangle$.}
}

%

\author{
Yoichi \textsc{Takeda,}\altaffilmark{1}
Dong-Il \textsc{Kang,}\altaffilmark{2}
Inwoo \textsc{Han,}\altaffilmark{3}
Byeong-Cheol \textsc{Lee,}\altaffilmark{3,4}
and 
Kang-Min \textsc{Kim}\altaffilmark{3}
}

\altaffiltext{1}{National Astronomical Observatory, 2-21-1 Osawa, 
Mitaka, Tokyo 181-8588}
\email{takeda.yoichi@nao.ac.jp}
\altaffiltext{2}{Gyeongsangnamdo Institute of Science Education,\\
75-18 Gajinri, Jinsungmyeon, Jinju, Gyeongnam 660-851, Korea}
\email{kangdongil@gmail.com}
\altaffiltext{3}{Korea Astronomy and Space Science Institute,
61-1 Whaam-dong, Youseong-gu, Taejon 305-348, Korea}
\email{iwhan@kasi.re.kr, bclee@boao.re.kr, kmkim@boao.re.kr}
\altaffiltext{4}{Department of Astronomy and Atmospheric Sciences,\\ 
Kyungpook National University, Daegu 702-701, Korea} 
%


\maketitle

\begin{abstract}
An extensive non-LTE abundance analysis based on Na~{\sc i} 5890/5896 
doublet lines was carried out for a large unbiased sample of $\sim 120$ 
A-type main-sequence stars (including 23 Hyades stars) covering 
a wide $v_{\rm e}\sin i$ range of $\sim$~10--300~km~s$^{-1}$, 
with an aim to examine whether the Na abundances in such A dwarfs 
can be reliably established from these strong Na~{\sc i} D lines.
The resulting abundances ([Na/H]$_{58}$), which were obtained by 
applying the $T_{\rm eff}$-dependent microturbulent velocities of
$\xi \sim$~2--4 km~s$^{-1}$ with a peak at $T_{\rm eff} \sim 8000$~K 
(typical for A stars), turned out generally negative with a large
diversity (from $\sim -1$ to $\sim 0$), while showing a sign 
of $v_{\rm e}\sin i$-dependence (decreasing toward higher 
rotation). However, the reality of this apparently subsolar trend 
is very questionable, since these [Na/H]$_{58}$ are systematically 
lower by $\sim$~0.3--0.6~dex than more reliable [Na/H]$_{61}$ 
(derived from weak Na~{\sc i} 6154/6161 lines for sharp-line stars). 
Considering the large $\xi$-sensitivity of the abundances derived from 
these saturated Na~{\sc i} D lines, we regard that 
[Na/H]$_{58}$ must have been erroneously underestimated,
suspecting that the conventional $\xi$ values are improperly too 
large at least for such strong high-forming Na~{\sc i} 5890/5896 lines,
presumably due to the depth-dependence of $\xi$ decreasing with height.
The nature of atmospheric turbulent velocity field 
in mid-to-late A stars would have to be more investigated  
before we can determine reliable sodium abundances from these
strong resonance D lines.
\end{abstract}

%


\section{Introduction}

In spite of the long history and substantial amount of spectroscopic
studies of A-type stars on the upper main sequence, we do not know 
yet much about their photospheric abundances of sodium.
The primary reason for this paucity may be attributed to the fact
that available Na lines are quite limited in such comparatively
hot early-type stars, because Na atoms are easily ionized to make
Na~{\sc i} lines quickly fade out as $T_{\rm eff}$ becomes higher 
(reflecting the characteristic of alkali elements having one 
valence electron weakly bound) while lines of Na~{\sc ii} (closed
shell) are hopeless to detect.

As a matter of fact, many of the past spectroscopic studies of A dwarfs
which reported Na abundances have placed emphasis on late A-type or Am 
stars of comparatively lower $T_{\rm eff}$ (typically at $\ltsim 8500$~K) 
where subordinate Na~{\sc i} lines such as those at 5683/5688 or 
6154/6161~$\rm\AA$ are still measurable (e.g., Lane \& Lester 1987;
Varenne \& Monier 1999;  Gebran et al. 2008; Gebran \& Monier 2008;
Fossati et al. 2007, 2008). Meanwhile, early-A stars have been rarely
investigated in this respect; actually, challenges of Na abundance
determinations were done mostly for very bright stars such as
Vega (cf. Takeda 2008 and references therein) or Sirius
(cf. Kohl 1964; Takeda \& Takada-Hidai 1994), while trials did not
turn out successful  when improper lines were used 
(e.g., Hill \& Landstreet 1993).

Besides, another problem is the bias toward sharp-line stars of slow rotation, 
since the Na~{\sc i} lines at 5683/5688 or 6154/6161~$\rm\AA$ 
which have been normally used as mentioned above, tend to be washed out 
in rapid rotators because of their weakness. This is especially the case
for early A stars, even though nowadays analyses invoking an efficient 
spectrum synthesis technique have succeeded in establishing the Na 
abundances of rapidly rotating late-A or Am stars (see the literature 
this decade mentioned above).

Now, it appears for us necessary to try understanding the nature of 
photospheric Na abundances in ``general'' A-type stars.
For example, a special attention has recently been paid to
sodium in relation to a special group of weak-lined A-type stars 
($\lambda$~Bootis stars), which spread in a wide $T_{\rm eff}$ range 
(from early to late A-type) and generally rotate rapidly. 
That is, their photospheric abundances of Na, showing
a large diversity from supersolar to subsolar unlike other metals
being generally deficient, appear to correlate with the Na composition
of the ambient interstellar matter, which indicates that accretion of
interstellar gas onto a star may have played some role in building
up the photospheric chemical peculiarity (cf. Kamp \& Paunzen 2002;
Paunzen et al. 2002). If such an external mechanism is involved, 
a question naturally arises whether such an effect is limited only to
a group of rapid rotators or it can also influence other normal A-type 
stars. In this respect, it is important and worthwhile to explore the 
Na abundances of an unbiased sample of early-to-late A-type stars 
in a large range of $v_{\rm e}\sin i$, which has never been challenged
to our knowledge. 
  
If we are to contend with this task, there is no other way than to
invoke the Na~{\sc i} resonance lines at 5890/5896~$\rm\AA$ 
(D$_{1}$ and D$_{2}$) in the orange region of stellar spectra,
which are so strong as to be visible in spectra of any A-type stars 
no matter high $T_{\rm eff}$ or $v_{\rm e}\sin i$ is.
However, it is not necessarily easy to use these D lines for Na abundance
determinations. Apart from the practical complexities in terms of
superficial spectrum appearance (e.g., necessity of removing nuisance 
telluric lines of water vapor existing in this wavelength region, 
need of applying a spectrum synthesis technique in case of rapid 
rotators where doublet lines are merged with each other), an important 
point is that these lines suffer a strong non-LTE effect, which has 
to be taken into account by all means. For this reason, several old
studies which treated Na~{\sc i} 5890/5896 lines in LTE are hardly 
reliable.

And yet, researches in this field are still insufficient. To our 
knowledge, only a few studies on the non-LTE effect of neutral Na 
lines applicable to A-type stars have been done so far: \\
--- Takeda and Takada-Hidai (1994) carried out a non-LTE analysis 
by using published equivalent width ($EW$) data of various sodium 
lines for Sirius (A1~V) as a by-product study of A--F supergiants. \\
--- Based on extensive statistical equilibrium calculations, Mashonkina, 
Shimanski\v{\i}, and Sakhibullin (2000) computed non-LTE corrections 
of various Na~{\sc i} spectral lines for a wide range of atmospheric 
parameters (4000~K~$\le T_{\rm eff} \le 12500$~K, $0.0 \le \log g \le 4.5$,
and $-4.0 \le$~[X/H]~$\le 0.5$). \\
--- Andrievsky et al. (2002) determined Na abundances of twelve 
$\lambda$~Bootis candidates (including rapid rotators) from Na~{\sc i} 
D$_{1}$ and D$_{2}$ lines by taking into account the non-LTE effect.\\
--- An intensive Na abundance study of Vega (A0~V), while taking account 
of the non-LTE as well as the gravity-darkening effect, has recently 
been carried out by Takeda (2008). 

In the latest work of Takeda (2008), a remarkable consistency
was obtained between the Na abundances of Vega derived from 8 lines 
(at 5683/5688, 5890/5896, 6154/6161, 8183/8195~$\rm\AA$) showing widely 
different non-LTE corrections. Encouraged by this success of our non-LTE 
calculations, we decided to conduct an extensive Na abundance study
on a large unbiased sample of A-type main-sequence stars based on 
non-LTE analyses of Na~{\sc i} 5890/5896 lines, with an intention of
clarifying the behavior of photospheric sodium abundances as well as 
their dependence (if any) upon stellar parameters such as $T_{\rm eff}$ 
or $v_{\rm e}\sin i$, since high-dispersion spectra for $\sim$~120 A 
dwarfs just suitable for this purpose were timely available to us. 
This was the original motivation of the present investigation.

In the course of our study, however, we began to realize that
our results appear rather questionable and can not be simply taken at 
face values, which indicates that the procedure of our analyses
needs to be reexamined, especially for late- to mid-A stars where
the microturbulent velocity is generally considered to be enhanced. More 
specifically, we have arrived at a conclusion that appropriately choosing
the turbulent velocity parameter (and correctly understanding the 
atmospheric velocity field) is essential for establishing reliable 
Na abundances from the strong high-forming Na~{\sc i} resonance lines, 
which does not seem to be an easy task at all. Thus, the purpose of 
this paper is to describe this outturn in detail.

\section{Observational Data}

The targets of this study are 122 A-type stars on the upper
main sequence, which is an extended sample of 46 stars studied
in Takeda et al. (2008; hereinafter referred to as Paper I). 
The basic data of these object are given in table 1.
(See also the electronic table~E for more detailed information.) 
Our sample includes not only stars classified as normal,
but also a number of Am stars (as in Paper I), three $\lambda$~Boo-type 
stars ($\lambda$~Boo, 29~Cyg, $\pi^{1}$~Ori), and one Ap star of 
Cr-type ($\epsilon$~UMa). Our sample also contains 23 A-type stars 
belonging to Hyades cluster, which are important in discussing
the nature/origin of abundance peculiarities because they
must have had almost the same composition when they were born.
These 122 program stars are plotted on the $\log L$ vs. $\log T_{\rm eff}$ 
diagram in figure 1, where theoretical evolutionary tracks 
corresponding to different stellar masses are also depicted.
We can see from this figure that the masses of our sample stars
are in the range between $\sim 1.5 M_{\odot}$ and $\sim 3 M_{\odot}$.

\setcounter{figure}{0}
\begin{figure}
  \begin{center}
    \FigureFile(70mm,70mm){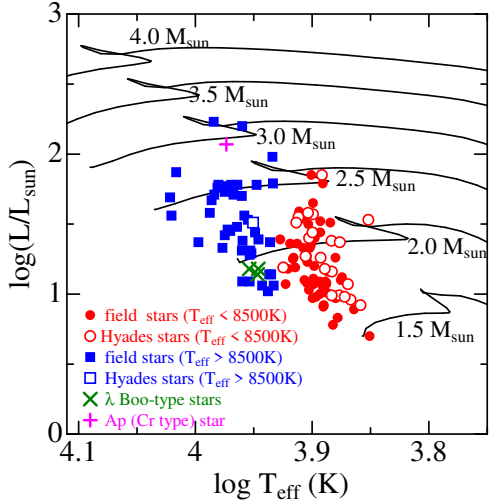}
  \end{center}
\caption{Plots of 122 program stars on the theoretical HR diagram
($\log (L/L_{\odot})$ vs. $\log T_{\rm eff}$), where the bolometric 
luminosity ($L$) was evaluated from the apparent visual magnitude with 
the help of Hipparcos parallax (ESA 1997) and Flower's (1996) bolometric
correction. Filled circles (red) --- field stars with $T_{\rm eff} < 8500$~K,
open circles (red) --- Hyades stars with $T_{\rm eff} < 8500$~K, 
filled squares (blue) --- field stars with $T_{\rm eff} > 8500$~K,
open squares (blue) --- Hyades stars with $T_{\rm eff} > 8500$~K,
St. Andrew's crosses ($\times$, green) --- $\lambda$ Boo stars,
and Greek crosses ($+$, pink) --- Ap star (Cr-type).
Theoretical evolutionary tracks corresponding to 
the solar metallicity computed by Girardi et al. (2000) for 
six different initial masses are also depicted for comparison.
}
\end{figure}

The observations were carried out on 2008 January 14--16,\footnote{
These 2008 January data were already used for the analysis of Paper I.
The observations and reductions for the 2008 September and 
2009 January data were done in the same manner.}
2008 September 6--9, and 2009 January 7--8 by using BOES (Bohyunsan 
Observatory Echelle Spectrograph) attached to the 1.8 m reflector 
at Bohyunsan Optical Astronomy Observatory. 
Using 2k$\times$4k CCD (pixel size of 15~$\mu$m~$\times$~15~$\mu$m), 
this echelle spectrograph enabled us to obtain spectra of wide 
wavelength coverage (from $\sim$~3700~$\rm\AA$ to 
$\sim$~10000~$\rm\AA$) at a time.
We used 200~$\mu$m fiber corresponding to the resolving power 
of $R \simeq 45000$. The integrated exposure time 
for each star was typically $\sim$~10--20~min on the average.
The reduction of the echelle spectra (bias subtraction, flat 
fielding, spectrum extraction, wavelength calibration, and 
continuum normalization) was carried out with the software 
developed by Kang et al. (2006). For most of the targets, we 
could accomplish sufficiently high S/N ratio of several hundreds 
at the most sensitive orange--red region. 


\section{Fundamental Parameters}

\subsection{Atmospheric Parameters}

As done in Paper I, The effective temperature ($T_{\rm eff}$) 
and the surface gravity ($\log g$) of each program star were 
determined from the colors of Str\"{o}mgren's $uvby\beta$ 
photometric system with the help of the {\tt uvbybetanew}\footnote{
$\langle$http://www.astro.le.ac.uk/\~{}rn38/uvbybeta.html$\rangle$.}
program (Napiwotzki et al. 1993).
For 12 stars, for which $uvby\beta$ data were not available, 
we evaluated $T_{\rm eff}$ from their $B-V$ color by using 
the $B-V$ vs. $T_{\rm eff}$ relation (for $\log g = 4.0$ and 
the solar metallicity) computed by Kurucz (1993), 
and $\log g$ from $L$ (luminosity), $M$ (mass) 
(estimated from the position on the theoretical HR diagram; 
cf. figure 1), and $T_{\rm eff}$. 
Regarding the microturbulence, we adopted the analytical 
$T_{\rm eff}$-dependent relation derived in Paper I,
\begin{equation}
\xi = 4.0 \exp\{- [\log (T_{\rm eff}/8000)/A]^{2}\} \\
\end{equation}
(where $A \equiv [\log (10000/8000)]/\sqrt{\ln 2}$),
which roughly represents the observed distribution of $\xi$
with probable uncertainties of $\pm 30\%$ (cf. figure 2b in Paper I).
The final values of $T_{\rm eff}$, $\log g$, and $\xi$ are 
summarized in table 1.

The model atmosphere for each star was then constructed
by two-dimensionally interpolating Kurucz's (1993) ATLAS9 
model grid in terms of $T_{\rm eff}$ and $\log g$, where
we exclusively applied the solar-metallicity models 
as in Paper I.

\subsection{Rotational Velocity and Abundances of Five Elements}

As in Paper I, we determined the projected rotational velocity 
($v_{\rm e}\sin i$) and the abundances of five elements 
(O, Si, Ca, Fe, and Ba)\footnote{
For the special sharp-line cases of 13 stars where 
Na~{\sc i} 6154/6161 lines are detectable, 
Na abundances were also determined in this fitting analysis
(by using Kurucz \& Bell's 1995 $gf$ values),
in addition to these 5 species.} by applying the synthetic-fitting 
technique to the spectrum portion of 6140--6170~$\rm\AA$
(cf. section 4 in Paper I for the details of the procedure).
Figure 2 demonstrates how the theoretical and observed spectra
match each other with the finally converged solutions, and
the resulting $v_{\rm e}\sin i$ as well as the differential 
abundances of these elements relative to Procyon\footnote{
As in Paper I, we adopted Procyon as the standard reference star,
which has essentially the same abundances as the Sun. See 
subsection IV-c in Paper I.}
[X/H] ($\equiv A_{\rm star} - A_{\rm Procyon}$) are given in 
table 1 as well as in electronic table E.\footnote{Since all 
46 stars studied in Paper I are included in the present sample 
of 122 stars, we redetermined their solutions in exactly the 
same way as for the newly 
added 76 stars. While the new and old results are practically 
equivalent, notable changes have resulted for some exceptional 
cases. This must have been caused by the delicate difference in 
the boundaries of the working spectrum region; i.e., we carefully 
adjusted the boundaries this time for each star by eye-inspection 
(so that they fall on the pseudo-continuum window), whereas 
we fixed them for all stars in the analysis of Paper I.}

\section{Sodium Abundance Determination}

\subsection{Removal of Telluric Lines} 

Since the region of the Na~{\sc i} 5890/5896 doublet lines
is more or less contaminated by telluric water vapor lines,
we had to remove them first by divining the raw spectrum
of each star by a relevant spectrum of a rapid rotator 
(with $v_{\rm e}\sin i$ values typically of 
$\sim$~200--300~km~s$^{-1}$) by using the IRAF task {\tt telluric}. 
(This reference spectrum had been appropriately ``re-''normalized 
prior to division so that its continuum level becomes cleanly flat.)
Two demonstrative examples (sharp- as well as broad-line cases)
of this elimination process are depicted in figure 3,
where we can recognize that it turned out successful.
Actually, in almost all cases of our 122 targets, the telluric 
features could be satisfactorily cleared away by this kind
of procedure (cf. figure 4).

\begin{figure}
\setcounter{figure}{2}
  \begin{center}
    \FigureFile(70mm,70mm){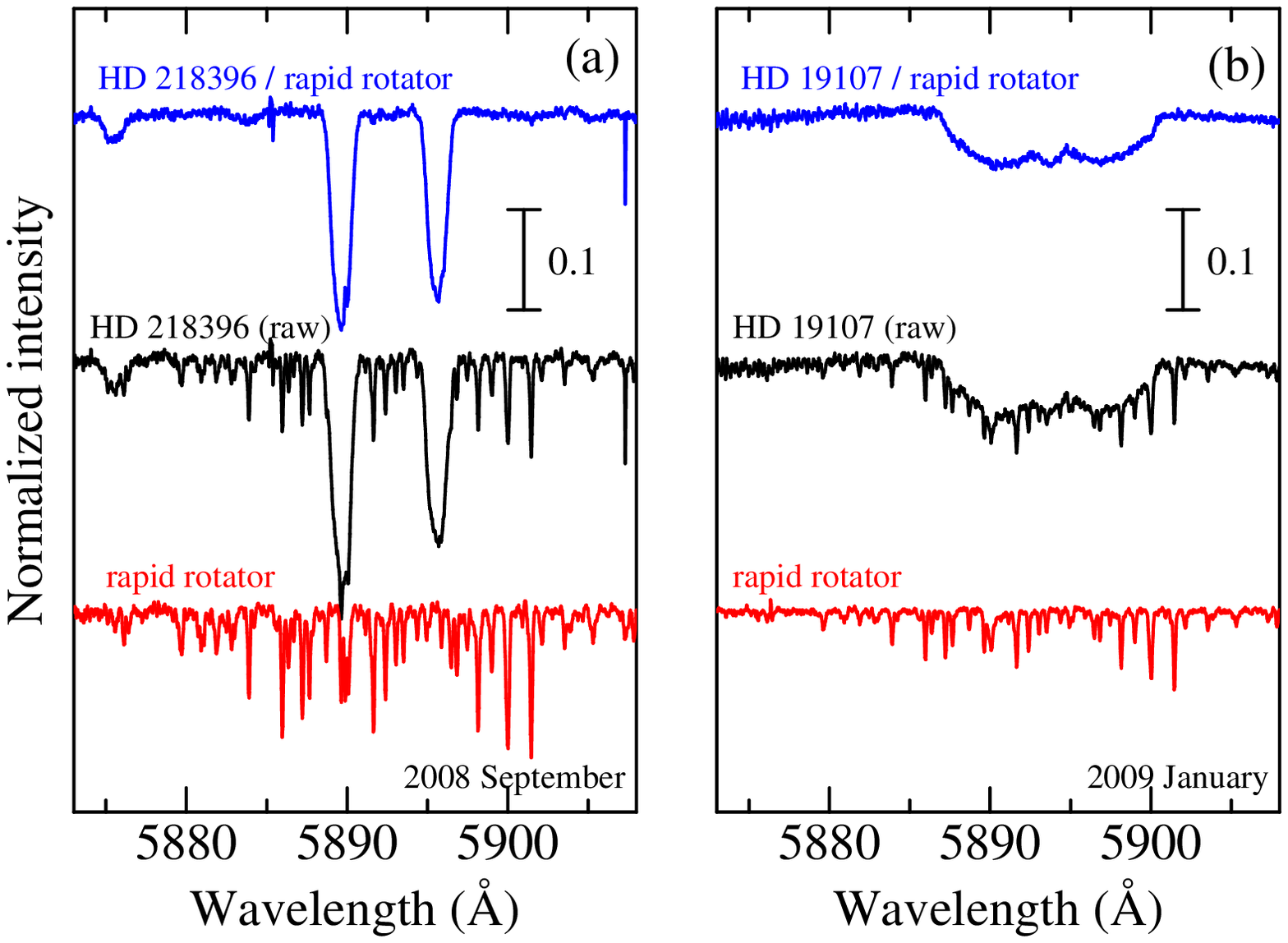}
  \end{center}
\caption{Examples of how the telluric lines (mostly due to
H$_{2}$O vapor) are removed in the Na~{\sc i} 5890/5896 region. 
Dividing the raw stellar spectrum (middle, black) by the spectrum 
of a rapid rotator (bottom, red) results in the final 
spectrum (upper, blue). The left (a) and right (b) panel 
show the typical sharp-line case (HD~218396) and the broad-line 
case (HD~19107), respectively. No Doppler correction is applied 
to the wavelength scale of these spectra. Note that the effect of
these telluric lines is less significant in the dry winter season
than in the wet summer season.
}
\end{figure}

\subsection{Non-LTE Synthetic Spectrum Fitting}

The non-LTE statistical-equilibrium calculations for neutral sodium 
were implemented for a grid of 44 models resulting from combinations 
of 11 $T_{\rm eff}$ values (7000, 7500, 8000, 8500, 9000, 9500, 10000,
10500, 11000, 11500, 12000~K) and 4 $\log g$ values (3.0, 3.5, 4.0, 4.5), 
so that we can obtain the depth-dependent non-LTE departure coefficients
for any star by interpolating (or extrapolating) this grid. 
See Takeda et al. (2003) for the computational details.

Now that the non-LTE departure coefficients relevant for each star 
are available, with which the non-LTE theoretical spectrum
of Na~{\sc i} 5890/5896 lines can be computed, we carried out
a spectrum-synthesis analysis by applying Takeda's (1995a) 
automatic-fitting procedure to the 5880--5905~$\rm\AA$ region 
while regarding $A_{\rm Na}$ (as well as $v_{\rm e}\sin i$ and 
radial velocity) as an adjustable parameter
to be established.\footnote{The abundances of Fe ($A_{\rm Fe}$) or 
Ni ($A_{\rm Ni}$) were also treated as variables in several special 
cases where features of these lines were appreciable.} 
The adopted atomic data of the relevant Na~{\sc i} lines are
presented in table 2.
How the theoretical spectrum for the converged solutions fits well 
with the observed spectrum is displayed in figure 4.

\subsection{Inverse Evaluation of Equivalent Widths}

Abundance determination based on the synthetic spectrum fitting is 
a powerful approach, since it allows reliable results even in 
cases of very broad lines where the Na~{\sc i} doublet lines
are severely merged and empirical measurements of individual 
equivalent widths ($EW$s) are hardly possible. 
However, this method is not necessarily useful when one wants 
to study the abundance sensitivity to changing the atmospheric 
parameters or the assumptions used in its derivation (i.e., it is 
very tedious to repeat the fitting process again and again for 
different set of atmospheric parameters). In this respect, $EW$ values 
are favorable, since they are much easier to handle. Hence, we computed 
the equivalent widths of the Na~{\sc i} 5890 ($EW_{5890}$) and 
Na~{\sc i} 5896 ($EW_{5896}$) ``inversely'' from the abundance 
(resulting from non-LTE spectrum synthesis) and the adopted atmospheric 
model/parameters. 

\subsection{Abundance Analysis with EW$_{5890}$ and EW$_{5896}$}

Based on such evaluated $EW_{5890}$ and $EW_{5896}$, we carried out 
a full-fledged abundance analysis while deriving the non-LTE (as well as 
LTE) Na abundances with the atomic parameters given in table 2. 
We used Kurucz's (1993) WIDTH9 program for this purpose, which had 
been considerably modified in various respects (especially to 
include the effect of departure from LTE).

Actually, since our preliminary study had revealed the tendency of subsolar 
Na abundance, we prepared two sets of non-LTE departure coefficients 
(grid of 44 models) corresponding to two different choices of input 
Na abundances ([Na/H] = 0.0 and [Na/H] = $-1.0$), and two kinds of 
non-LTE abundances ($A_{\rm Na}^{0}$ and $A_{\rm Na}^{-1}$) 
were obtained for a given $EW$ for each of the two sets. Then, these
$A_{\rm Na}^{0}$ and $A_{\rm Na}^{-1}$ were interpolated (or 
extrapolated) so that the final non-LTE solution ($A_{\rm Na}$) and 
the used departure coefficient becomes consistent with each other 
(cf. subsection 4.2 in Takeda \& Takada-Hidai 1994). In addition, 
we also derived the LTE abundance $A_{\rm Na}^{\rm LTE}$, from which
the non-LTE correction was derived as $\Delta^{\rm NLTE} \equiv
A_{\rm Na} - A_{\rm Na}^{\rm LTE}$. Finally, the differential Na 
abundance relative to Procyon was computed as 
[Na/H] $\equiv A_{\rm Na} - 6.35$, where $A_{\rm Na}^{\rm Procyon} = 6.35$ 
is the non-LTE Na abundance of Procyon (derived from the fitting
analysis in the 5880--5895~$\rm\AA$ region; cf. subsection 4.2), 
which is very close to the solar Na 
abundance of 6.33 (Anders \& Grevesse 1989).
The resulting values of $\langle$[Na/H]$\rangle$, 
$\langle EW \rangle$, and $\langle \Delta^{\rm NLTE} \rangle$
(averaged values for the two lines) for each of the program stars 
are summarized in table 1 (see electronic table E for the more detailed 
line-by-line results).

We also estimated the uncertainties in $A_{\rm Na}$
by repeating the analysis while perturbing
the standard values of the atmospheric parameters 
interchangeably by $\pm 300$~K in $T_{\rm eff}^{\rm std}$, 
$\pm 0.3$~dex in $\log g^{\rm std}$, and $\pm 30\%$ in 
$\xi^{\rm std}$ (which are the typical uncertainties of 
the parameters we adopted; cf. section IV-c in Paper I). 
Such evaluated abundance changes ($\Delta_{T+}$, $\Delta_{T-}$, 
$\Delta_{g+}$, $\Delta_{g-}$, $\Delta_{\xi +}$, 
and $\Delta_{\xi -}$) are given in electronic table E
(and also shown in figures 5d--f).

\setcounter{table}{1}
\begin{table}[h]
\scriptsize
\caption{Atomic data of Na~{\sc i} 5890 and 5896 lines.}
\begin{center}
\begin{tabular}
{crrrrrr}\hline \hline
RMT & $\lambda$ ($\rm\AA$) & $\chi$ (eV) & $\log gf$ & Gammar & Gammas &
Gammaw \\
\hline
1  & 5889.951 & 0.000 &   0.117  & 7.80 & $-5.64$ & $-7.67$ \\
1  & 5895.924 & 0.000 & $-0.184$ & 7.80 & $-5.64$ & $-7.67$ \\
\hline
\end{tabular}
\end{center}
\scriptsize
Note. \\
All data are were taken from Kurucz \& Bell's (1995) compilation.\\
Followed by first four self-explanatory columns,
damping parameters are given in the last three columns:\\
Gammar is the radiation damping width (s$^{-1}$), $\log\gamma_{\rm rad}$.\\
Gammas is the Stark damping width (s$^{-1}$) per electron density (cm$^{-3}$) 
at $10^{4}$ K, $\log(\gamma_{\rm e}/N_{\rm e})$.\\
Gammaw is the van der Waals damping width (s$^{-1}$) per hydrogen density 
(cm$^{-3}$) at $10^{4}$ K, $\log(\gamma_{\rm w}/N_{\rm H})$. \\
\end{table}

\subsection{Characteristics of the Results}

These quantities resulting from our analysis are plotted against 
$T_{\rm eff}$ in figure 5.
As we can see from figure 5a, the $\langle EW \rangle$'s of 
Na~{\sc i} 5890/5896 lines are progressively weakened as 
$T_{\rm eff}$ becomes higher, reflecting the strong temperature 
sensitivity of these resonance lines of neutral alkali atoms. 
Also, the extents of the (negative) non-LTE corrections are 
generally considerable and important, amounting from 0.2~dex to 
0.8~dex (figure 5b). While the corrections are 
almost constant in late-A stars ($T_{\rm eff} \ltsim 8500$~K) 
where lines are so strong to be saturated, they show a rather 
large scatter for early-A stars ($T_{\rm eff} \gtsim 8500$~K) 
depending on the abundance of each star.

\setcounter{figure}{4}
\begin{figure}
  \begin{center}
    \FigureFile(70mm,70mm){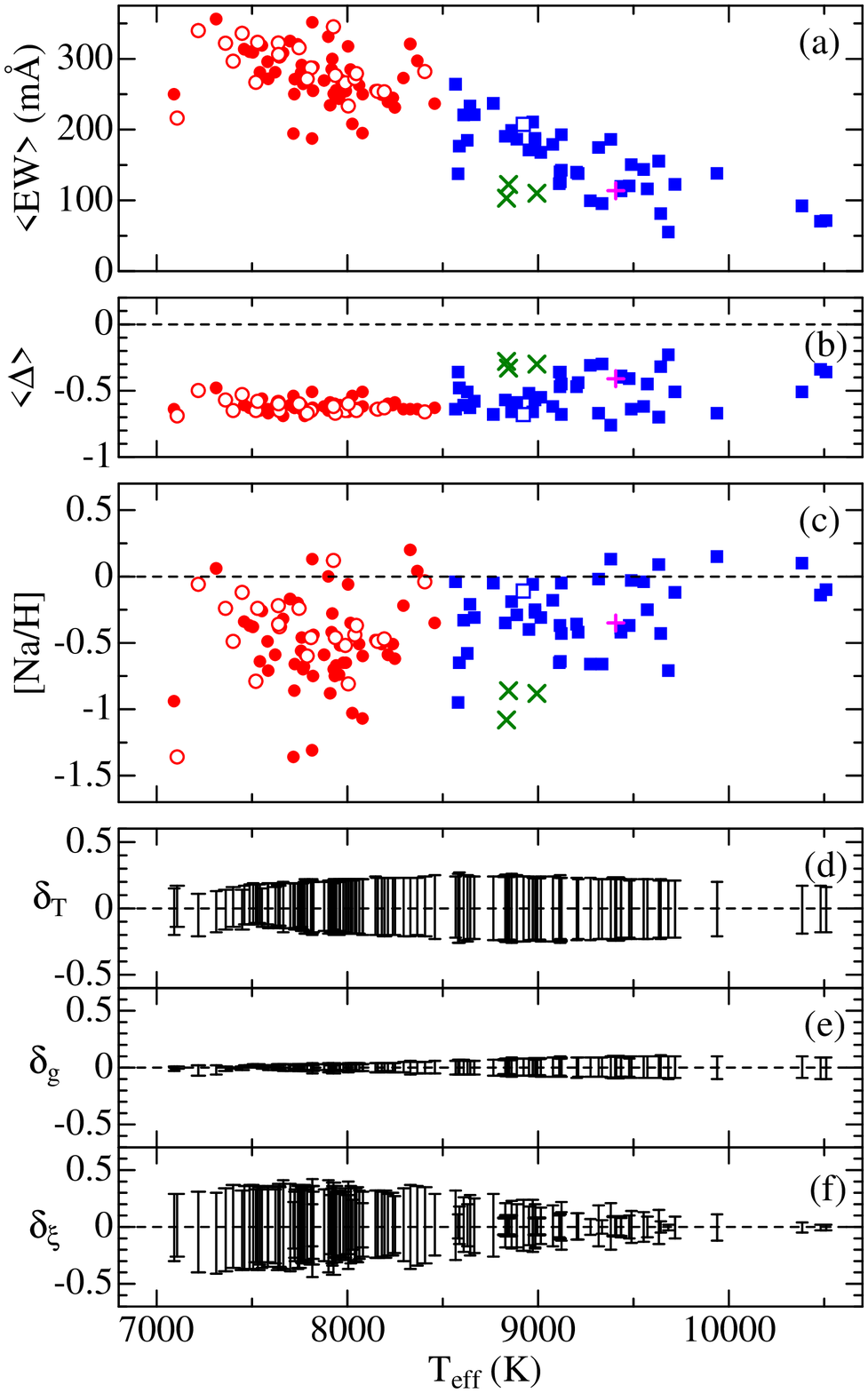}
  \end{center}
\caption{Sodium abundances derived from Na~{\sc i} 5890/5896 lines
and the related quantities plotted against $T_{\rm eff}$. (a) Mean 
equivalent width ($\langle EW \rangle$; average of $EW_{5890}$ and
$EW_{5896}$), (b) mean non-LTE correction ($\langle \Delta \rangle$; 
average of $\Delta_{5890}$ and $\Delta_{5896}$),
(c) [Na/H] (sodium abundance relative to the standard star
Procyon), (d) $\delta_{T+}$ and $\delta_{T-}$ (Na abundance 
variations in response to $T_{\rm eff}$ changes of
+300~K and $-300$~K), (e) $\delta_{g+}$ and $\delta_{g-}$ 
(Na abundance variations in response to $\log g$ changes 
of $+0.3$~dex and $-0.3$~dex), and (f) $\delta_{\xi +}$ and 
$\delta_{\xi -}$ (Na abundance variations in response to 
changing $\xi$ as $\xi \times 1.3$ and $\xi / 1.3$).
See the caption of figure 1 for the meanings of the symbols 
in (a)--(c). 
}
\end{figure}

The trend of [Na/H], manifesting itself in figure 5c, is
somewhat surprising. That is, [Na/H] exhibits a considerably
large diversity (from $\sim -1$ to $\sim 0$) at 
$T_{\rm eff} \sim$~8000--8500~K, which appears to gradually 
shrink toward a higher $T_{\rm eff}$
([Na/H] $\sim 0$ at $T_{\rm eff} \sim 10000$~K).
If we take this result at face value, we would have to
conclude that Na is apparently underabundant in the photosphere
of A-type main-sequence stars. However, these results should 
not be taken seriously, since there are good reasons
to believe that we have considerably underestimated the sodium 
abundances from these strong Na~{\sc i} 5890/5896 lines
presumably due to the adoption of a too large microturbulence 
($\xi$), as discussed in the next section.

Regarding the three $\lambda$ Boo stars included in our sample
($\lambda$~Boo, 29~Cyg, and $\pi^{1}$~Ori, which are of the 
classical $\lambda$~Boo type; cf. Baschek \& Searle 1969), 
they apparently show a markedly lower [Na/H] (cf. those denoted 
by ``+'' in figure 5c) compared to other normal A dwarfs 
in accordance with their metal-deficient nature ([Fe/H] from 
$\sim -0.7$ to $\sim -1.5$; cf. table 1). Since the [Na/H] values
of these three stars 
are quite similar to each other, we could not confirm the recently 
reported results of considerably divergent [Na/H] in $\lambda$ Boo
stars (e.g., Andrievsky et al. 2002; Paunzen et al. 2002),
as far as these objects are concerned. 

\section{Discussion}

\subsection{Comparison with the Abundances from 6154/6161 Lines}

The striking evidence for the considerable underestimation of 
the sodium abundances 
derived from Na~{\sc i} 5890/5896 lines ([Na/H]$_{58}$) is shown 
in figures 6a and b, where [Na/H]$_{58}$ values are compared with
[Na/H]$_{61}$ (Na abundance relative to Procyon derived from 
Na~{\sc i} 6154/6161 lines,\footnote{
In this case, the reference sodium abundance ($A_{\rm Procyon}$),
which was derived from the 6140--6170~$\rm\AA$ fitting (figure 2)
by using Takeda et al.'s (2005) Procyon spectrum, was 6.39.} 
which could be accomplished for 13 sharp-line stars;
cf. footnote 3 in subsection 3.2). As seen from these figures,
while a positive correlation surely exists between [Na/H]$_{58}$
and [Na/H]$_{61}$ (figure 6a), they show a marked discrepancy
amounting to $\sim$~0.3--0.6~dex ([Na/H]$_{61} >$ [Na/H]$_{58}$), 
which appears to be $T_{\rm eff}$-dependent
in the sense that the discordance tends to become milder toward a higher
$T_{\rm eff}$. Since the abundances derived from Na~{\sc i} 6154/6161 
lines are believed to be reliable, because they are too weak to
be influenced by any ``external'' parameters in the analysis
(e.g., a choice of microturbulence or how the non-LTE effect
is treated), we can not help considering that our [Na/H]$_{58}$
values must have been largely underestimated (by $\sim 0.5$~dex
at $T_{\rm eff} \sim 8000$~K and $\sim 0.3$~dex at 
$T_{\rm eff} \sim 9000$~K). If we take into account these 
systematic errors in figure 5c, [Na/H] values in A-type dwarfs 
would become nearly solar on the average irrespective of $T_{\rm eff}$
(though still with a considerable scatter).

\setcounter{figure}{5}
\begin{figure}
  \begin{center}
    \FigureFile(70mm,70mm){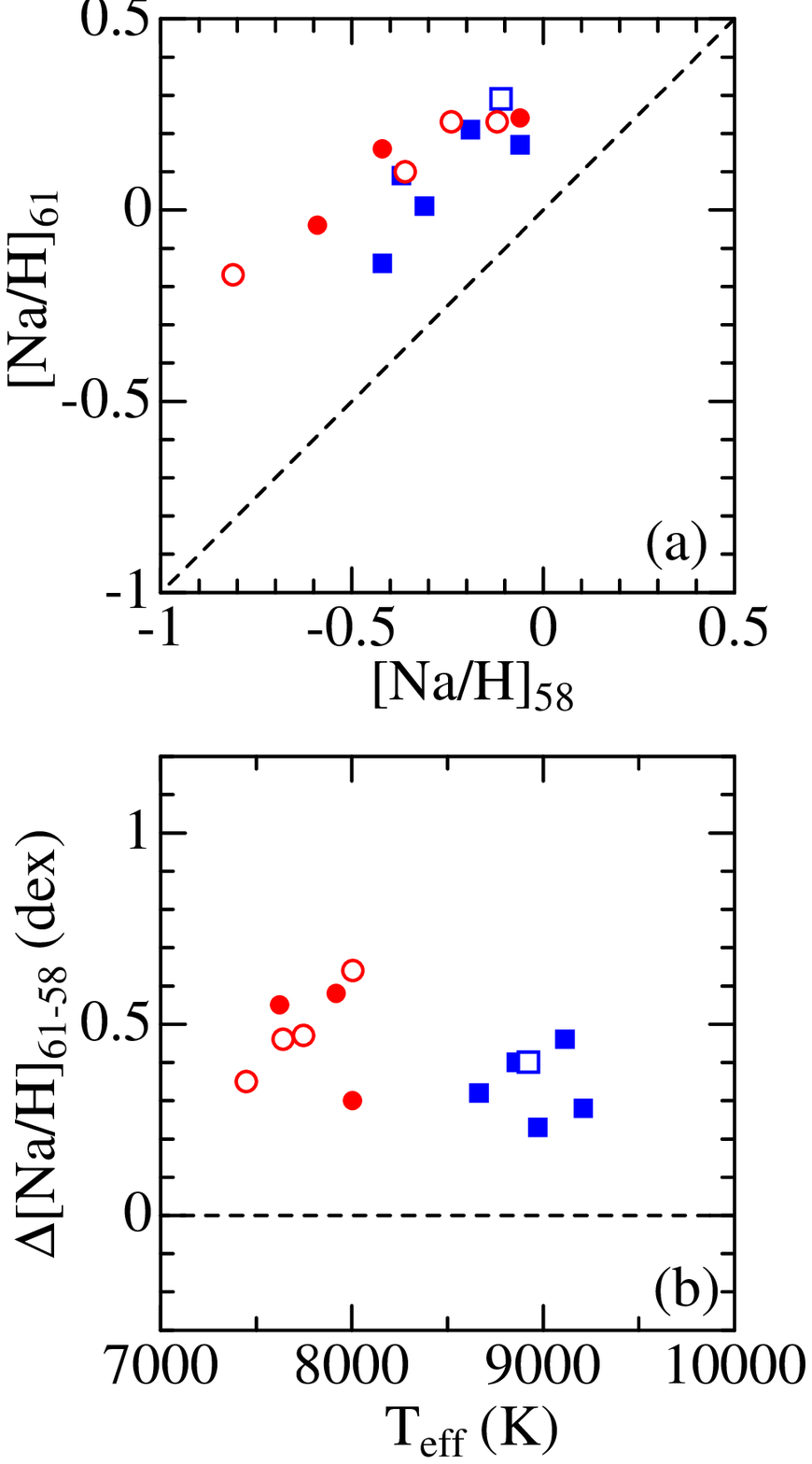}
  \end{center}
\caption{(a) Comparison of the sodium abundances derived from 
Na~{\sc i} 5890/5896 lines ([Na/H]$_{58}$; column 14 in table 1) 
with those determined from the 6140--6170~$\rm\AA$ fitting based 
on the weak Na~{\sc i} 6154/6161 lines ([Na/H]$_{61}$, only for 
13 stars; cf. column 17 in table 1). (b) Difference of 
[Na/H]$_{61}$~$-$~[Na/H]$_{58}$ plotted against $T_{\rm eff}$. 
See the caption of figure 1 for the meanings of the symbols.
}
\end{figure}

\subsection{Is Assigned Microturbulence Valid?}

Then, what has caused such systematic errors (i.e., underestimation) 
in our non-LTE sodium abundances derived from the strong 
Na~{\sc i} 5890/5896 lines? 
According to figures 5d--f, $T_{\rm eff}$ or $\xi$ may be
counted as important parameters in the sense that their errors 
might cause significant influences on [Na/H]. Among these two,
if we are to ascribe the cause of error solely to $T_{\rm eff}$,
it must be systematically too low by more than $\gtsim 500$~K 
which we consider unlikely. Instead, we believe that the
error must have stemmed from an improper choice of $\xi$,
since the $T_{\rm eff}$ region of $\sim 8000$~K where [Na/H] is 
most $\xi$-sensitive (figure 5f) coincides with that showing
particularly large (superficial) underabundance (cf. figure 5c). 
Namely, we seem to have
assigned improperly too large $\xi$ values in determining 
$A_{\rm Na}$ from Na~{\sc i} 5890/5896 lines, which in consequence 
lead to a considerable underestimation of the abundance, reflecting
the strong $\xi$-sensitivity of $A_{\rm Na}$ derived from these 
strongly saturated lines.  

Then, was equation (1) for $\xi$ as a function of $T_{\rm eff}$ 
we invoked for evaluating $\xi$ of each star inappropriate?
While the qualitative trend of the $\xi$ vs. $T_{\rm eff}$ relation 
represented by equation (1) (cf. figure 2b in Paper I) is 
certainly correct according to a number of studies so far (see below), 
we admittedly are not very confident about whether the maximum 
$\xi$ value ($\xi_{\rm max}$) of $\sim 4$~km~s$^{-1}$ (attained at 
$T_{\rm eff} \sim 8000$~K) implied by equation (1) is really 
adequate in the quantitative sense, since previous studies have 
reported rather different $\xi_{\rm max}$ results from each other; i.e., 
$\sim$~9~km~s$^{-1}$ (Baschek \& Reimers 1969),
$\sim$~7~km~s$^{-1}$ (Smith 1971),
$\sim$~3~km~s$^{-1}$ (Coupry \& Burkhart 1992),
$\sim$~5~km~s$^{-1}$ (Takeda \& Sadakane 1997a),
$\sim$~5~km~s$^{-1}$ (Varenne \& Monier 1999),
$\sim$~3~km~s$^{-1}$ (Smalley 2004), and
$\sim$~3~km~s$^{-1}$ (Gebran \& Monier 2007).
Nevertheless, we can state by examining these results
that our choice of $\xi_{\rm max} \sim$~4~km~s$^{-1}$ is
surely reasonable, considering that all recent studies after 1990 
point to a value in the range of $\sim$~3--5~km~s$^{-1}$
(which is just within our estimated uncertainties of $\pm 30\%$). 
To say the least, a $\xi_{\rm max}$ value as low as $\sim 2$~km~s$^{-1}$,
which is required to raise [Na/H]$_{58}$ at $T_{\rm eff} \sim 8000$~K
by $\sim 0.6$~dex to make it consistent with [Na/H]$_{58}$ (cf.
figures 5c and 6b), is quite unlikely. 

\subsection{Depth-Dependence of $\xi$: Possible Key to the Solution}

We rather consider that the essential problem lies in the conventional 
assumption of depth-independent uniform microturbulence.
The Na~{\sc i} 5890/5896 resonance lines form at markedly higher 
layer compared to other metallic lines usually used for determining 
$\xi$,\footnote{For the present case of A-type stars, the typical 
mean line-formation depth $\langle \log \tau \rangle$ (in terms of 
the continuum optical depth at 5000~$\rm\AA$; cf. table 2 of 
Takeda \& Takada-Hidai 1994 for its definition) for the Na~{\sc i} 5890/5896 
lines is in the range of $\langle \log \tau \rangle \sim -3$ to $-2$, 
while that for the typical metallic lines (such as the Fe~{\sc ii} 
line at 6147.74~$\rm\AA$ or Ca~{\sc i} line at 6162.17~$\rm\AA$ 
conspicuously seen in the 6140--6170~$\rm\AA$ region; cf. figure 2)
is $\langle \log \tau \rangle \gtsim -1$.} because of their 
considerable strengths as well as their low-excitation nature favoring 
an environment of lower temperature. So, in case that $\xi$ appreciably 
decreases with height, it is natural that our applying the conventionally
established larger $\xi$ (e.g., from deep-forming metallic lines) 
to Na abundance determinations from high-forming Na~{\sc i} 
5890/5896 lines (for which smaller $\xi$ is relevant) must have 
lead to a serious underestimation of the abundances. 

While available observational studies on the depth-dependence of 
$\xi$ in the atmosphere of dwarf stars are only limited (especially, 
the relevant late-A dwarfs showing a peak in $\xi$ have barely 
been investigated in this respect), there is a suggestion that 
$\xi$ tends to decrease with an increase in the atmospheric height, 
especially for F--G dwarfs (including the Sun) where the turbulent 
velocity field is believed to originate from the convective (granular) 
motion.\footnote{On the contrary, in the case of evolved giants or 
supergiants of low gravity ($\log g \ltsim 3$), previous investigations 
suggested that $\xi$ tends to increase with height (e.g., Takeda 1992; 
Takeda \& Takada-Hidai 1994; Takeda \& Sadakane 1997b).}
For example, Takeda et al. (1996) showed in the analysis of
the strong solar K~{\sc i} 7699 line that the profile-based $\xi$  
(reflecting the $\xi$ at higher layer where the line core is formed)
is appreciably smaller than the conventional $\xi$ determined
from equivalent widths, indicating a decreasing $\xi$ with height
(see also Canfield \& Beckers 1976). Similarly, they also concluded 
from the analysis of many lines of different forming depths that
$\xi$ in the atmosphere of Procyon (F5~IV--V) decreases with
height (cf. Appendix 2 therein). Such a depth-dependent tendency 
of $\xi$ may be interpreted that the turbulence is associated with
convective overshooting in which the motion is decelerated 
toward upper layers. As a matter of fact, the ``macro''-turbulence
of larger scale also shows such a decreasing tendency with height
(cf. Takeda 1995b). 

Meanwhile, in the case of hotter early-A dwarfs 
($T_{\rm eff} \sim 10000$~K), where the origin of $\xi$ may be 
not so much due to convection (an effective convection zone is 
unlikely to exist) as a sound wave or pulsation as suggested 
by Gigas (1986), the situation is still rather controversial. 
While some investigators suggest (from the spectroscopic study 
in UV as well as visual region) that $\xi$ tends to ``decrease'' 
with height (for A0~V star Vega by Castelli \& Faraggiana 1979 
or for A0~IV star $\gamma$~Gem by Nishimura \& Sadakane 1994), 
other researchers report an opposite tendency of ``increasing'' 
$\xi$ with height (for A0~V star Vega by Gigas 1986 or for 
A2~IV star $o$~Peg by Zboril 1992).

Returning to late-A dwarfs ($T_{\rm eff} \sim 8000$~K) of our
primary concern, such an empirical study on the depth-dependence 
of $\xi$ seems to be lacking to our knowledge. However,
we may reasonably assume that the origin of $\xi$ is due to
convective overshooting as in F--G dwarfs, since the reason why
$\xi$ attains a maximum value in this temperature range is
considered be that the convection zone comes fairly close to 
the surface atmospheric layer, which means that decreasing $\xi$
with an increase in height would equally apply to this case.

Interestingly, Freytag and Steffen's (2004; cf. figure 2 therein) 
3D numerical simulations of convection in the atmosphere of late-A 
stars ($T_{\rm eff} = 8500$~K) shows that the r.m.s. dispersion of 
the vertical component of the turbulent velocity ($v_{\rm vert,rms}$)
steeply declines with height from $\sim$~4~km~s$^{-1}$ 
(at $z \simeq -0.5$~Mm) down to $\sim$~1~km~s$^{-1}$ 
(at $z \simeq 0.5$~Mm), followed by a small up and down 
(between $\sim$~0.5~km~s$^{-1}$ and $\sim$~2~km~s$^{-1}$) at 
$z \simeq$~1--2~Mm.
Since the difference of the geometrical height between the 
optically-thin layer (e.g., $\tau \sim 10^{-2}$ to $10^{-4}$) 
and the continuum-forming layer ($\tau \sim 1$) is on the order
of $\sim$~0.4--0.8~Mm in the atmosphere of A-type dwarfs, their 
simulation indicates that $\xi$ quickly decreases
from $\sim$~4~km~s$^{-1}$ to $\sim$~1~km~s$^{-1}$ over the whole 
range of the atmosphere relevant for line formation.
Such a theoretical prediction, suggesting a progressive attenuation 
of $\xi$ over the line-forming region with increasing height, 
favorably lends support for our interpretation. 

In order to assess the impact of this effect on the abundance
derived from Na~{\sc i} 5890/5896 lines, we performed a test 
calculation based on a simple model of depth-dependent $\xi$, 
decreasing linearly in terms of $\log \tau$ from 
4~km~s$^{-1}$ (at $\log\tau = 0$) to 1~km~s$^{-1}$ 
(at $\log\tau = \log\tau_{\rm s}$, where $\log\tau_{\rm s}$ 
is a free parameter controling the gradient of $\xi$); that is,  
$\xi$ = 4~km~s$^{-1}$ (at $\log\tau \ge 0$), 
$\xi =  4 - 3 \log\tau / \log \tau_{\rm s}$~km~s$^{-1}$
(at $0 \ge \log \tau \ge \log \tau_{\rm s}$), and 
$\xi$ = 1~km~s$^{-1}$ (at $\log\tau_{\rm s} \ge \log \tau$).
As an example, we selected the sharp-line star HD~40932 
($T_{\rm eff} = 8005$~K, $\log g = 3.93$, $\xi = 4.0$~km~s$^{-1}$),
for which [Na/H]$_{58}$ (= $-0.81$) and [Na/H]$_{61}$ (= $-0.17$) 
show a considerable disagreement (0.64~dex) with each other (cf. table 1).
We determined [Na/H]$_{58}$ with this microturbulence model
for three cases of $\log\tau_{\rm s}$ = $-4$, $-3$, and $-2$,
and found that the corresponding abundance changes (compared to the
standard value derived from the constant $\xi$ of 4.0~km~s$^{-1}$)
were +0.47, +0.62, and +0.81 (for the 5890 line) 
and +0.35, +0.48, and +0.66 (for the 5896 line), respectively.
These corrections are sufficient to raise [Na/H]$_{58}$ to remove 
the discrepancy of 0.64~dex between [Na/H]$_{58}$ and [Na/H]$_{61}$,
since the latter abundance derived from the weak Na~{\sc i} 6154/6161 
lines ($EW \sim 10$~m$\rm\AA$) are practically unaffected by any 
change of $\xi$. 

Consequently, we would conclude that 
the decreasing $\xi$ with the atmospheric height is the cause 
for the erroneous underestimation of [Na/H]$_{58}$.
If we are ever to derive the correct sodium abundance from Na~{\sc i} 
5890/5896 lines, a classical uniform (depth-independent) 
microturbulence is hardly applicable any more. It would be inevitable 
to take into account the realistic depth-dependent turbulent velocity 
field (preferably based on detailed 3D hydrodynamical simulations).

\subsection{Parameter Dependence of [Na/H]}

Finally, even though we were thus obliged to conclude that the [Na/H] 
results for 122 A-type stars we derived from Na~{\sc i} 5890/5896 lines 
were considerably underestimated and should not be taken at face 
values, it would still be meaningful to discuss their apparent 
dependence on other parameters.

The trend of underabundance in [Na/H] (most prominent for late-A 
stars of $T_{\rm eff} \sim 8000$K) appears to become progressively 
insignificant toward higher $T_{\rm eff}$ and eventually ends up 
with [Na/H]~$\sim 0$ at $T_{\rm eff} \sim 10000$~K (cf. figure 5c),
which  may be explained 
by (1) the lowered sensitivity of [Na/H] to $\xi$ because of the 
weakened (less saturated) $EW$, and (2) the decreasing tendency of 
$\xi$ with height may not be manifest any more because of the 
practical disappearance of the convection zone (cf. 3rd paragraph 
in the previous subsection). 

Regarding the dependence of rotation, [Na/H] appears to show
a systematic decrease with an increase in $v_{\rm e}\sin i$ 
as displayed in figure 7a. Since [Fe/H] also shows such a tendency
(Am-like overabundance exhibited by slowly-rotating stars
disappears in rapid rotators; cf. figure 8e or Paper I), a positive 
correlation is apparently seen between [Na/H] and [Fe/H] (figure 7b). 
While this might indicate that an anomaly of Na abundance is built-up
in a manner similar to that of Fe (i.e., advent of Na excess
in slow rotators), we are not sure whether this is really the case,
since Na (in a marked contrast to Fe) is not considered to show 
an appreciable abundance peculiarity according to the recent 
theory of atomic diffusion (e.g., Richer et al. 2000). 
Rather, we would consider a possibility that the discrepancy of $\xi$
between high and lower layer (i.e., decreasing rate of $\xi$ with 
atmospheric height) may become more exaggerated  (and the erroneous 
underabundance due to the use of improper $\xi$ being more prominent)
as a star rotates higher, though whether such an effect is theoretically 
reasonable is yet to be investigated. 

\setcounter{figure}{6}
\begin{figure}
  \begin{center}
    \FigureFile(70mm,70mm){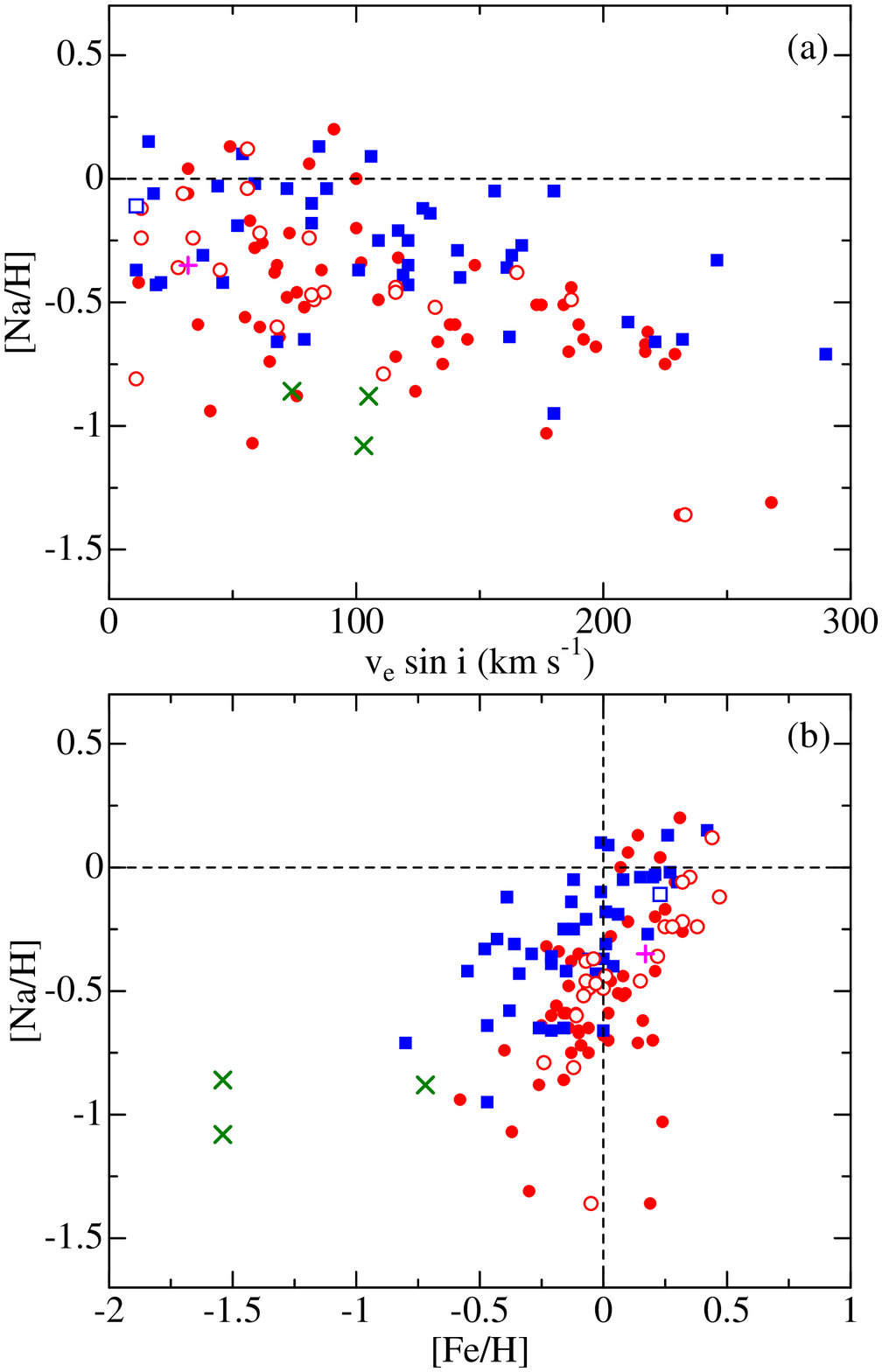}
  \end{center}
\caption{Sodium abundances ([Na/H]) plotted against (a) 
the projected rotational velocity ($v_{\rm e} \sin i$)
and (b) the Fe abundance ([Fe/H]). See the caption of figure 1 
for the meanings of the symbols.
}
\end{figure}

\section{Conclusion}

Based on the high-dispersion spectrum data obtained with the 1.8~m
reflector and BOES spectrograph at Bohyunsan Optical Astronomy 
Observatory, we determined the sodium abundances of 122 slowly-
as well as rapidly-rotating A-type stars from the Na~{\sc i} 
5890/5896 doublet lines, while including the non-LTE effect based on
detailed statistical-equilibrium calculations, in order to 
check whether the Na abundances in such A dwarfs can be reliably 
established from such strong resonance lines, and (if possible) 
to investigate the behavior of [Na/H] for such dwarf stars on the 
upper main sequence in the $T_{\rm eff}$ range of $\sim$~7000--10000~K. 
Regarding the important parameter of atmospheric microturbulence 
dispersion ($\xi$), we adopted an empirically derived analytical 
formula as a function of $T_{\rm eff}$, which presents $\xi$ 
in the range of $\sim$~2--4 km~s$^{-1}$ with a peak at 
$T_{\rm eff} \sim 8000$~K 

We found that the resulting abundances ([Na/H]) relative to the 
standard star Procyon (equivalent to the Sun) turned out generally 
negative with a large diversity (from $\sim -1$ to $\sim 0$), 
while showing a sign of $v_{\rm e}\sin i$-dependence (decreasing 
toward higher rotation). 

However, the credibility of these apparent trends is very 
questionable, since such abundances derived from strong
$\xi$-sensitive Na~{\sc i}~5890/5896 lines are appreciably 
lower by $\sim$~0.3--0.6~dex than the abundances based on the 
weak Na~{\sc i} 6154/6161 lines (which are much more reliable
but usable only for sharp-line stars). We thus concluded
that our Na abundances from these Na~{\sc i} resonance D lines
must have been erroneously underestimated.

We suspect that this failure may be attributed to the possible
depth-dependence (i.e., a decreasing tendency with height) of $\xi$ 
expected for turbulent velocity field of convective overshooting
origin, which implies that the conventionally determined $\xi$ 
values (e.g., from the equivalent widths of Fe lines) are 
improperly too large to apply to such strong high-forming 
Na~{\sc i} 5890/5896 lines.

Although it turned out unfortunately unsuccessful to establish 
the sodium abundances of A dwarfs from the Na~{\sc i} D lines, 
in spite of our original intention, one should not consider that 
these strong lines are practically useless. On the contrary, 
they may be used as an important touchstone for any sophisticated 
theory or 3D hydrodynamical simulation of the complicated velocity 
field in the atmosphere of A-type stars, possibly with the requirement
that [Na/H] should have a value close to $\sim 0$ for normal
stars. We would have to wait until we have a clear understanding
about this velocity field, before being able to determine
reliable Na abundances therefrom.

\bigskip

This research has made use of the SIMBAD database, operated by
CDS, Strasbourg, France. 
I. Han acknowledges the financial support for this study by KICOS through 
Korea--Ukraine joint research grant (grant  07-179).
B.-C. Lee acknowledges the Astrophysical Research Center for the Structure 
and Evolution of the Cosmos  (ARSEC, Sejong University) of the 
Korea Science and Engineering Foundation (KOSEF) through the Science
Research Center (SRC) program.

\appendix

\section*{Rotation-Dependence of [X/H] Revisited}

As described in subsection 3.2, we determined $v_{\rm e}\sin i$,
[O/H], [Si/H], [Ca/H], [Ba/H], and [Fe/H] for 122 program stars
by way of the synthetic spectrum fitting in the 6140--6170~$\rm\AA$ 
region as in Paper I. Since the number of stars has been considerably
increased (compared to the study in Paper I) and the present sample
includes 23 A-type stars of the Hyades cluster (which can be used
as a desirable probe of ``posteriori'' acquired peculiarity since 
their initial composition must have been almost uniform), 
it may be worthwhile to reexamine the $v_{\rm e}\sin i$-dependence 
of [X/H] based on the present results.

In figures 8a--e are shown the [X/H] values for each 5 elements 
plotted against $v_{\rm e}\sin i$. Comparing these figures with 
figures 10b, c, d, f, and g in Paper I, we could confirm that
the present results are naturally consistent with what was concluded
in Paper I, while some trends/problems have become newly apparent 
thanks to the considerably enlarged sample:\\
--- Am-like abundance peculiarities (i.e., underabundances of O and Ca,
overabundances of Fe and Ba) depends upon the rotational velocity
(i.e., decrease with increasing $v_{\rm e}\sin i$)
at 0~km~s$^{-1} \ltsim v_{\rm e}\sin i \ltsim 100$~km~s$^{-1}$.\\
--- At $v_{\rm e}\sin i \gtsim 100$~km~s$^{-1}$, [O/H], [Ca/H],
and [Fe/H] are almost constant at $\sim 0$.\\
--- Regarding Hyades A stars, the behaviors of [O/H] and [Fe/H]
(along with the mutual anti-correlation) are almost in agreement 
with the conclusion of Takeda and Sadakane (1997a), though
the $v_{\rm e}\sin i$-dependence appears to be limited at 
$v_{\rm e}\sin i \ltsim 100$~km~s$^{-1}$ rather than the wider
$v_{\rm e}\sin i$ range of $\ltsim 200$~km~s$^{-1}$ suggested
in that paper.\\
--- The large scatter in [Si/H] (cf. figure 8) would simply reflect 
that these results are not sufficiently reliable (especially for rapid
rotators or higher-$T_{\rm eff}$ stars) because they are based on
comparative weak Si~{\sc i} lines . Actually, 
[Si/H] could not be determined for a number of stars (cf. table 1).\\
--- That [Ba/H] of high rotators 
($v_{\rm e}\sin i \gtsim 100$~km~s$^{-1}$) becomes appreciably 
subsolar (figure 8d; also recognizable in figure 10f
of Paper I) is presumably unreal and should not be earnestly taken.
We suspect that this may be the same phenomenon as seen in 
[Na/H] of rapid rotators (figure 7a), to which a hypothetical
explanation described at the end of subsection 5.4 might apply.

\setcounter{figure}{7}
\begin{figure}
  \begin{center}
    \FigureFile(70mm,70mm){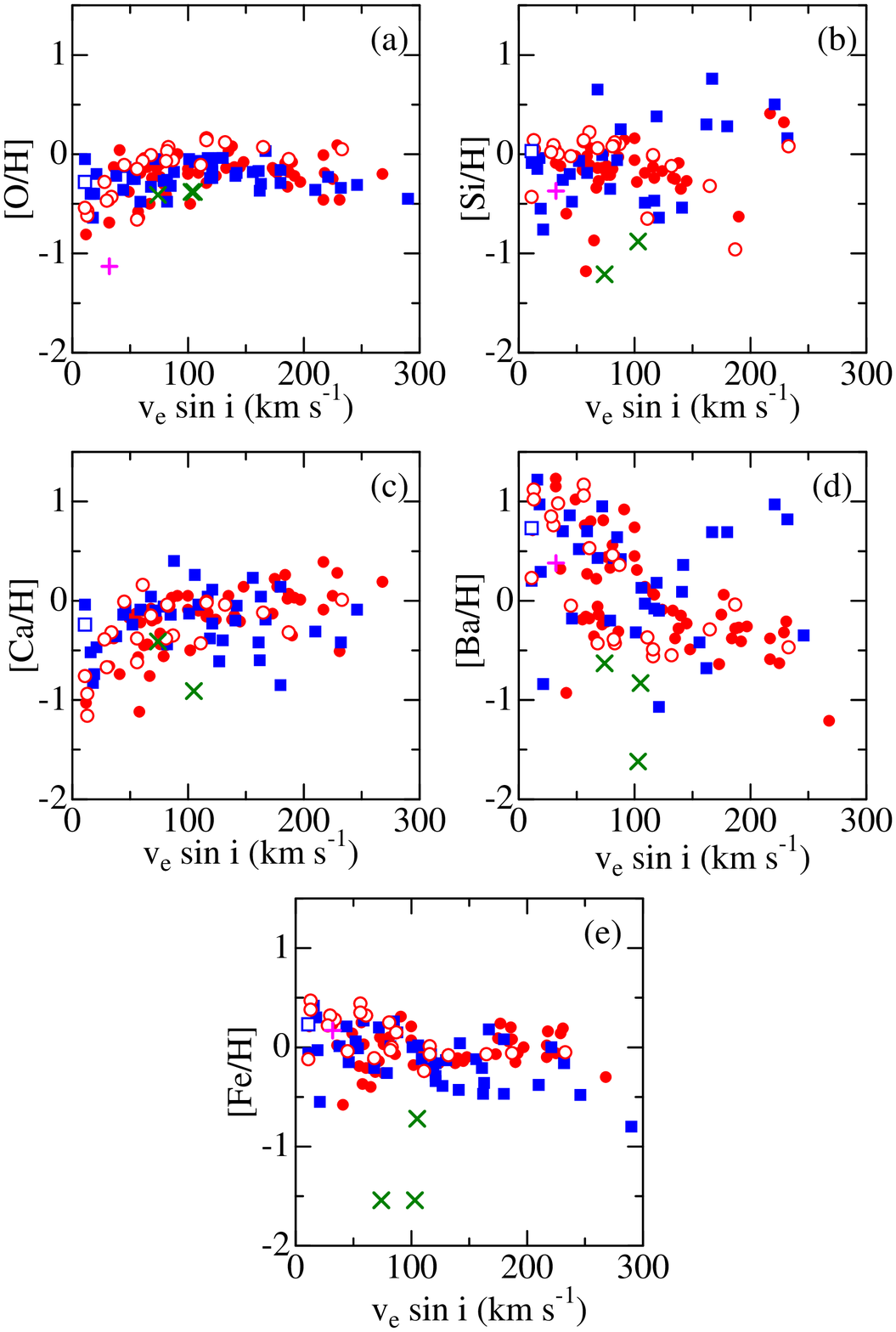}
  \end{center}
\caption{[X/H] values plotted against $v_{\rm e} \sin i$,
based on the results obtained from the spectrum fitting
analysis in the 6140--6170~$\rm\AA$ region: 
(a) [O/H], (b) [Si/H], (c) [Ca/H], (d) [Ba/H], and (e) [Fe/H]. 
See the caption of figure 1 for the meanings of the symbols.
}
\end{figure}

\clearpage
\twocolumn

\setcounter{figure}{1}
\begin{figure}
  \begin{center}
    \FigureFile(170mm,240mm){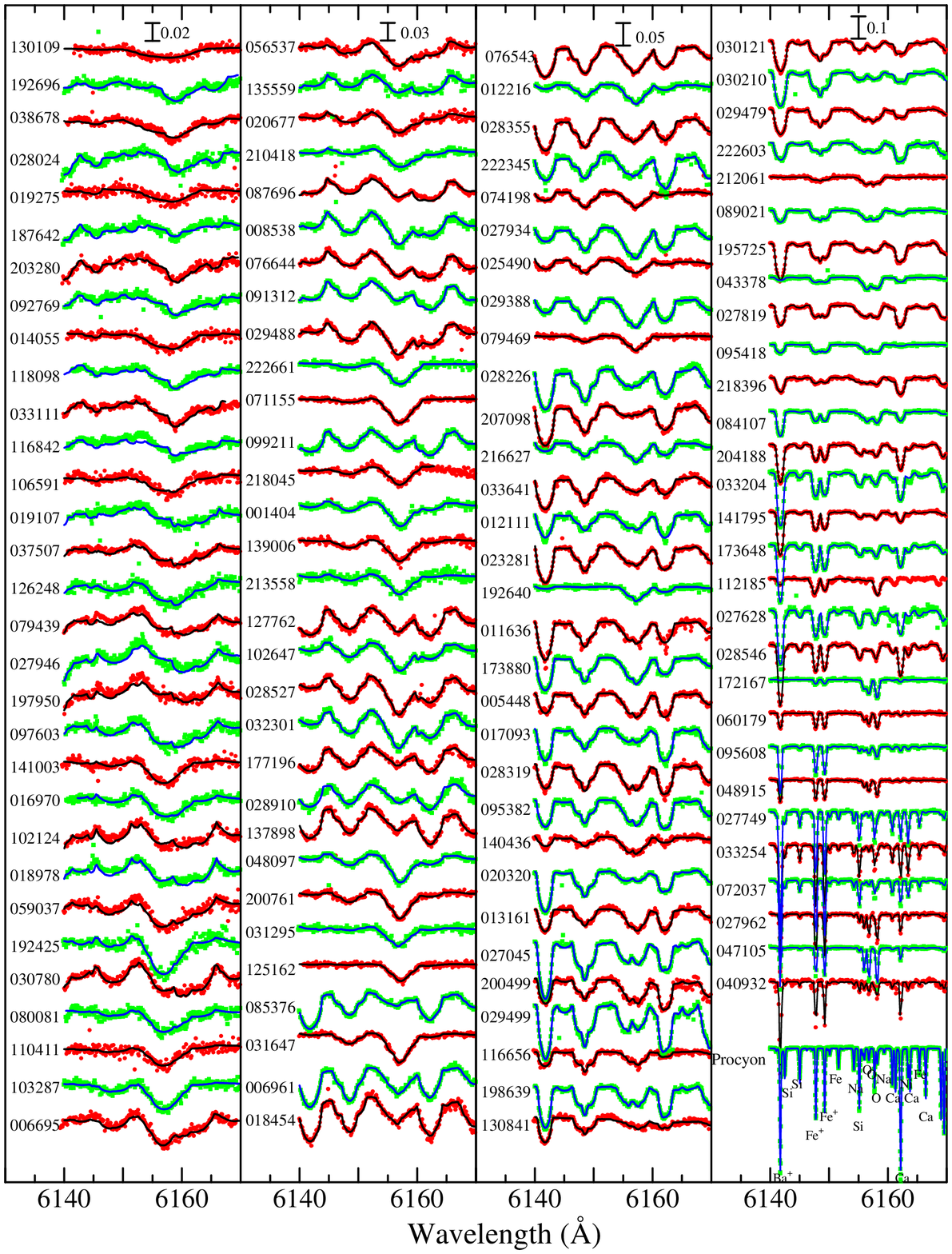}
  \end{center}
\caption{Synthetic spectrum fitting at the 6150 region 
(6140--6170~$\rm\AA$) for determining the projected rotational 
velocity ($v_{\rm e} \sin i$) along with the abundances of 
O, Si, Ca, Fe, and Ba.
The best-fit theoretical spectra are shown by solid lines, 
while the observed data are plotted by symbols.  
In each panel, the spectra are arranged in the descending order 
of $v_{\rm e} \sin i$ as in table 1, and an appropriate offset is 
applied to each spectrum (indicated by the HD number) relative to 
the adjacent one. The case of Procyon (standard star) is 
displayed at the bottom of the rightmost panel.
}
\end{figure}

\clearpage

\setcounter{figure}{3}
\begin{figure}
  \begin{center}
    \FigureFile(170mm,240mm){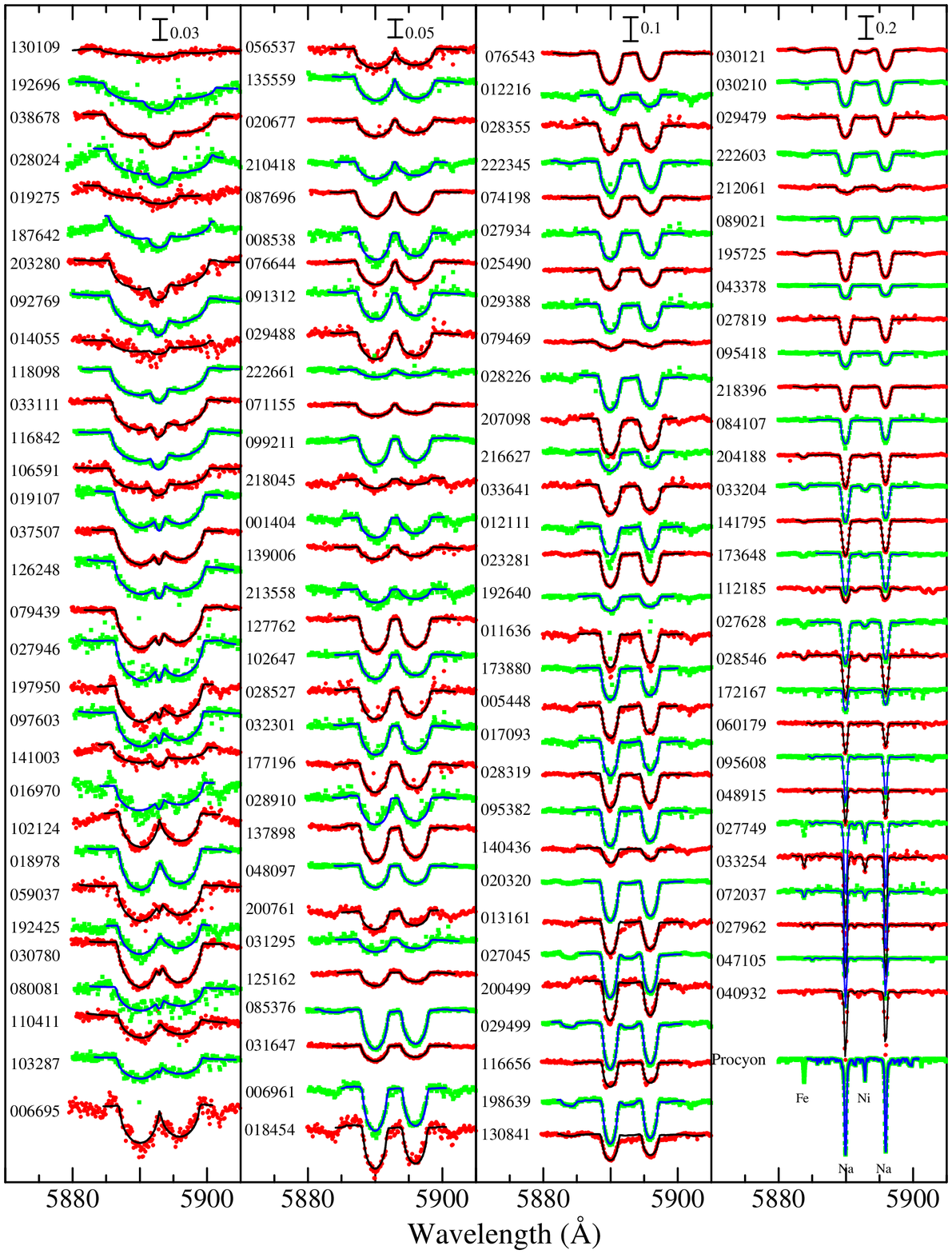}
  \end{center}
\caption{Synthetic spectrum fitting at the Na~{\sc i} 5890/5896 
region for establishing the abundance of sodium. Otherwise,
the same as in figure 2.
}
\end{figure}

\clearpage
\setcounter{table}{0}
\scriptsize
\renewcommand{\arraystretch}{0.8}
\setlength{\tabcolsep}{3pt}
\begin{longtable}{r@{ }r@{ }c@{}c@{}r@{ }c@{ }c@{}c@{ }c@{ }c@{ }c@{ }c@{ }c@{ }r@{ }c@{ }r l}
\caption{Basic stellar data and the results of the analysis.}
\hline\hline
HD\# & HR\# & Name & Sp.type & $T_{\rm eff}$ & $\log g$ & $\xi$ & 
$v_{\rm e}\sin i$ & O/H & Si/H & Ca/H & Fe/H & Ba/H & 
Na/H & $\langle EW \rangle$  & $\langle \Delta \rangle$ & Remark \\
\hline
\endhead
\hline
\endfoot
\hline
\multicolumn{17}{l}{\hbox to 0pt{\parbox{150mm}{\footnotesize
In columns 1 through 7 are given the HD number, HR number, star name
(with constellation), spectral type,
effective temperature (in K), logarithmic surface gravity (in  cm~s$^{-2}$), 
and microturbulent velocity (in km~s$^{-1}$). Columns 8 through 13 show 
the results determined from 6140--6170 region fitting: the projected 
rotational velocity (in km~s$^{-1}$), [O/H], [Si/H], [Ca/H], [Fe/H], 
and [Ba/H]. The final sodium abundance 
$\langle$[Na/H]$\rangle^{\rm NLTE}_{5890/5896}$ 
(mean of [Na/H]$_{5890}^{\rm NLTE}$ and [Na/H]$_{5896}^{\rm NLTE}$), mean 
equivalent width (average of $EW_{5890}$ and $EW_{5896}$; in m$\rm\AA$), 
mean non-LTE correction 
(average of $\Delta_{5890}$ and $\Delta_{5896}$; in dex) are given 
in columns 14--16, respectively. 
All abundance results ([X/H]; in dex) are the differential values 
relative to Procyon. 
The 122 stars are arranged in the descending order of $v_{\rm e} \sin i$,
to make it consistent with figures 2 and 4. 
In addition, the sodium abundances ([Na/H]$_{61}$) derived from the 
6140--6170 region fitting based on Na~{\sc i} 6154/6161 lines
(determinable only for 13 sharp-lined stars) are given in the final 
column 17 (values in parentheses), where 23 Hyades A-type stars are 
also denoted with ``H''. \\ \\
}}}
\endlastfoot
\hline
130109&5511&109 Vir&A0V         &   9683&  3.68&  2.4&   290& $-$0.45& $\cdots$& $\cdots$& $-$0.80& $\cdots$& $-$0.71&    55& $-$0.23&  \\
192696&7740&33 Cyg&A3IV-Vn     &   7815&  3.49&  4.0&   268& $-$0.20& $\cdots$& +0.19& $-$0.30& $-$1.21& $-$1.31&   187& $-$0.51&  \\
 38678&1998&$\zeta$ Lep&A2Vann      &   8610&  3.96&  3.7&   246& $-$0.31& $\cdots$& $-$0.09& $-$0.48& $-$0.35& $-$0.33&   221& $-$0.61&  \\
 28024&1392&$\upsilon$ Tau&A8Vn        &   7107&  3.20&  3.3&   233& +0.05& +0.08& +0.01& $-$0.05& $-$0.47& $-$1.36&   216& $-$0.69& H \\
 19275& 932&         &A2Vnn       &   9111&  4.12&  3.2&   232& $-$0.34& +0.16& $-$0.42& $-$0.16& +0.82& $-$0.65&   124& $-$0.36&  \\
187642&7557&$\alpha$ Aql&A7IV-V      &   7717&  4.00&  3.9&   231& $-$0.46& $\cdots$& $-$0.51& +0.19& $-$0.21& $-$1.36&   194& $-$0.54&  \\
203280&8162&$\alpha$ Cep&A7IV-V      &   7585&  3.73&  3.8&   229& +0.09& +0.32& +0.28& +0.14& $-$0.32& $-$0.71&   272& $-$0.67&  \\
 92769&4189&40 LMi&A4Vn        &   7820&  4.01&  4.0&   225& $-$0.25& $\cdots$& +0.05& $-$0.06& $-$0.63& $-$0.75&   255& $-$0.63&  \\
 14055& 664&$\gamma$ Tri&A1Vnn       &   9335&  3.98&  2.9&   221& $-$0.23& +0.50& $\cdots$& +0.00& +0.97& $-$0.66&    95& $-$0.30&  \\
118098&5107&$\zeta$ Vir&A3V         &   8249&  4.02&  4.0&   218& $-$0.19& $\cdots$& $\cdots$& +0.16& $\cdots$& $-$0.62&   231& $-$0.59&  \\
 33111&1666&$\beta$ Eri&A3IIIvar    &   7928&  3.59&  4.0&   217& $-$0.01& +0.41& +0.39& +0.02& $-$0.38& $-$0.70&   250& $-$0.64&  \\
116842&5062&80 UMa&A5V SB      &   7942&  4.02&  4.0&   217& $-$0.46& $\cdots$& $-$0.09& $-$0.10& $-$0.59& $-$0.67&   256& $-$0.62&  \\
106591&4660&$\delta$ UMa&A3Vvar      &   8629&  3.85&  3.7&   210& $-$0.36& $\cdots$& $-$0.31& $-$0.38& $\cdots$& $-$0.58&   185& $-$0.51&  \\
 19107& 925&$\rho^{3}$ Eri&A8V         &   7772&  3.96&  4.0&   197& $-$0.28& $\cdots$& +0.01& +0.00& $-$0.26& $-$0.68&   266& $-$0.64&  \\
 37507&1937&49 Ori&A4V         &   7979&  3.82&  4.0&   192& $-$0.22& $\cdots$& +0.03& $-$0.06& $-$0.41& $-$0.65&   253& $-$0.63&  \\
126248&5392&         &A5V         &   8212&  4.02&  4.0&   190& $-$0.08& $-$0.63& $-$0.35& $-$0.15& $-$0.27& $-$0.59&   239& $-$0.60&  \\
 79439&3662&18 UMa&A5V         &   7822&  4.03&  4.0&   187& $-$0.18& $\cdots$& +0.07& +0.08& $-$0.28& $-$0.44&   288& $-$0.64&  \\
 27946&1388&$\kappa^{2}$ Tau&A7V         &   7401&  3.84&  3.7&   187& $-$0.05& $-$0.96& $-$0.32& $-$0.06& $-$0.04& $-$0.49&   297& $-$0.65& H \\
197950&7945&4 Cep&A8V         &   7768&  4.08&  4.0&   186& $-$0.33& $\cdots$& +0.02& +0.20& $-$0.03& $-$0.70&   265& $-$0.63&  \\
 97603&4357&$\delta$ Leo&A4V         &   8180&  3.90&  4.0&   184& $-$0.09& $\cdots$& +0.26& +0.06& $-$0.38& $-$0.51&   249& $-$0.63&  \\
141003&5867&$\beta$ Ser&A3V         &   8580&  3.56&  3.7&   180& $-$0.29& $\cdots$& $-$0.85& $-$0.47& $\cdots$& $-$0.95&   137& $-$0.36&  \\
 16970& 804&$\gamma$ Cet&A3V         &   9122&  4.05&  3.2&   180& $-$0.16& +0.28& +0.14& +0.08& +0.69& $-$0.05&   193& $-$0.68&  \\
102124&4515&$\xi$ Vir&A4V         &   8026&  4.09&  4.0&   177& $-$0.19& $\cdots$& +0.14& +0.24& +0.06& $-$1.03&   208& $-$0.54&  \\
 18978& 919&$\tau^{3}$ Eri&A4V         &   8062&  4.03&  4.0&   175& $-$0.17& $\cdots$& +0.22& +0.09& $-$0.14& $-$0.51&   263& $-$0.63&  \\
 59037&2857&64 Gem&A4V         &   8238&  3.99&  4.0&   173& $-$0.14& $\cdots$& $-$0.13& $-$0.07& $-$0.64& $-$0.51&   245& $-$0.61&  \\
192425&7724&$\rho$ Aql&A2V         &   8984&  4.21&  3.3&   167& +0.03& +0.76& $-$0.19& +0.18& +0.69& $-$0.27&   185& $-$0.57&  \\
 30780&1547&97 Tau&A7IV-V      &   7644&  3.87&  3.9&   165& +0.07& $-$0.32& $-$0.12& $-$0.07& $-$0.29& $-$0.38&   304& $-$0.65& H \\
 80081&3690&38 Lyn&A1V         &   9014&  3.82&  3.3&   163& $-$0.30& $\cdots$& +0.04& $-$0.36& $\cdots$& $-$0.31&   168& $-$0.55&  \\
110411&4828&$\rho$ Vir&A0V         &   9117&  4.22&  3.2&   162& $-$0.37& +0.30& $-$0.60& $-$0.47& $-$0.68& $-$0.64&   126& $-$0.36&  \\
103287&4554&$\gamma$ UMa&A0V SB      &   9202&  3.79&  3.0&   161& $-$0.17& $\cdots$& $-$0.42& $-$0.21& $\cdots$& $-$0.36&   140& $-$0.47&  \\
  6695& 328&$\psi^{2}$ Psc&A3V         &   8765&  4.13&  3.6&   156& $-$0.18& $\cdots$& +0.23& $-$0.12& $-$0.42& $-$0.05&   237& $-$0.68&  \\
 56537&2763&$\lambda$ Gem&A3V...      &   8458&  3.90&  3.8&   148& $-$0.08& $\cdots$& +0.14& $-$0.10& $-$0.49& $-$0.35&   236& $-$0.63&  \\
135559&5679&4 Ser&A4V         &   7992&  4.14&  4.0&   145& $-$0.13& $-$0.27& $-$0.21& $-$0.14& $-$0.23& $-$0.65&   254& $-$0.61&  \\
 20677&1002&32 Per&A3V         &   8952&  4.08&  3.3&   142& $-$0.19& $\cdots$& $-$0.05& +0.04& +0.36& $-$0.40&   171& $-$0.52&  \\
210418&8450&$\theta$ Peg&A2V         &   8888&  3.82&  3.4&   141& $-$0.22& $-$0.54& $-$0.20& $-$0.43& +0.09& $-$0.29&   186& $-$0.59&  \\
 87696&3974&21 LMi&A7V         &   7878&  4.13&  4.0&   140& $-$0.13& $-$0.35& $-$0.15& $-$0.11& $-$0.15& $-$0.59&   269& $-$0.62&  \\
  8538& 403&$\delta$ Cas&A5Vv SB     &   7776&  3.41&  4.0&   138& +0.08& $-$0.09& $-$0.19& $-$0.16& $-$0.28& $-$0.59&   273& $-$0.69&  \\
 76644&3569&$\iota$ UMa&A7IV        &   7934&  4.22&  4.0&   135& +0.03& $-$0.25& +0.05& $-$0.13& $-$0.38& $-$0.75&   248& $-$0.60&  \\
 91312&4132&         &A7IV        &   7724&  4.08&  3.9&   133& $-$0.14& $-$0.24& $-$0.03& $-$0.10& $-$0.10& $-$0.66&   271& $-$0.63&  \\
 29488&1479&$\sigma^{2}$ Tau&A5Vn        &   7990&  3.82&  4.0&   132& +0.12& $-$0.12& $-$0.04& $-$0.08& $-$0.55& $-$0.52&   266& $-$0.65& H \\
222661&8988&$\omega^{2}$ Aqr&B9V         &  10481&  4.28&  1.5&   130& $-$0.04& $\cdots$& $-$0.40& $-$0.13& $\cdots$& $-$0.14&    70& $-$0.34&  \\
 71155&3314&         &A0V         &   9718&  4.11&  2.4&   127& $-$0.04& $\cdots$& $-$0.61& $-$0.39& $\cdots$& $-$0.12&   123& $-$0.51&  \\
 99211&4405&$\gamma$ Crt&A9V         &   7722&  3.95&  3.9&   124& $-$0.22& $-$0.17& $-$0.19& $-$0.16& $-$0.09& $-$0.86&   250& $-$0.63&  \\
218045&8781&$\alpha$ Peg&B9.5III     &   9643&  3.52&  2.5&   121& $-$0.24& $\cdots$& $\cdots$& $-$0.34& $\cdots$& $-$0.43&    81& $-$0.32&  \\
  1404&  68&$\sigma$ And&A2V         &   8828&  4.00&  3.5&   121& $-$0.13& $-$0.64& $-$0.23& $-$0.29& $-$0.10& $-$0.35&   190& $-$0.57&  \\
139006&5793&$\alpha$ CrB&A0V         &   9573&  3.87&  2.5&   121& $-$0.18& $\cdots$& +0.11& $-$0.16& $-$1.07& $-$0.25&   116& $-$0.45&  \\
213558&8585&$\alpha$ Lac&A1V         &   9434&  4.14&  2.7&   119& $-$0.24& +0.38& $-$0.38& $-$0.21& +0.18& $-$0.39&   120& $-$0.40&  \\
127762&5435&$\gamma$ Boo&A7IIIvar    &   7663&  3.59&  3.9&   117& $-$0.10& $-$0.24& $-$0.12& $-$0.23& +0.06& $-$0.32&   308& $-$0.69&  \\
102647&4534&$\beta$ Leo&A3Vvar      &   8643&  4.17&  3.7&   117& $-$0.04& $-$0.47& +0.04& $-$0.07& $-$0.08& $-$0.21&   233& $-$0.63&  \\
 28527&1427&         &A6IV        &   8039&  3.99&  4.0&   116& +0.16& $-$0.02& $-$0.01& +0.01& $-$0.56& $-$0.44&   272& $-$0.64& H \\
 32301&1620&$\iota$ Tau&A7V         &   7937&  3.74&  4.0&   116& +0.14& $-$0.01& $-$0.02& $-$0.07& $-$0.49& $-$0.46&   277& $-$0.67& H \\
177196&7215&16 Lyr&A7V         &   7940&  4.10&  4.0&   116& $-$0.29& $-$0.13& $-$0.16& $-$0.09& +0.07& $-$0.72&   250& $-$0.61&  \\
 28910&1444&$\rho$ Tau&A8V         &   7520&  3.97&  3.8&   111& $-$0.11& $-$0.65& $-$0.43& $-$0.24& $-$0.37& $-$0.79&   267& $-$0.65& H \\
137898&5746&10 Ser&A8IV        &   7582&  3.97&  3.8&   109& $-$0.19& $-$0.19& $-$0.10& $-$0.05& +0.14& $-$0.49&   296& $-$0.64&  \\
 48097&2466&26 Gem&A2V         &   8984&  4.23&  3.3&   109& $-$0.07& $-$0.49& $-$0.04& $-$0.12& $-$0.03& $-$0.25&   187& $-$0.57&  \\
200761&8075&$\theta$ Cap&A1V         &   9633&  4.11&  2.5&   106& $-$0.08& $\cdots$& +0.26& +0.02& +0.13& +0.09&   155& $-$0.70&  \\
 31295&1570&$\pi^{1}$ Ori&A0V         &   8993&  4.11&  3.3&   105& $-$0.38& $\cdots$& $-$0.91& $-$0.72& $-$0.83& $-$0.88&   110& $-$0.30&  \\
125162&5351&$\lambda$ Boo&A0sh        &   8834&  4.08&  3.5&   103& $-$0.38& $-$0.88& $\cdots$& $-$1.54& $-$1.62& $-$1.08&   103& $-$0.28&  \\
 85376&3900&22 Leo&A5IV        &   7459&  3.98&  3.7&   102& $-$0.50& $-$0.28& $-$0.50& $-$0.18& +0.31& $-$0.34&   314& $-$0.61&  \\
 31647&1592&$\omega$ Aur&A1V         &   9478&  4.27&  2.7&   101& $-$0.05& $\cdots$& $-$0.13& +0.00& $-$0.32& $-$0.37&   120& $-$0.41&  \\
  6961& 343&$\theta$ Cas&A7Vvar      &   7900&  3.81&  4.0&   100& $-$0.15& $-$0.06& $-$0.09& +0.07& +0.45& +0.00&   331& $-$0.65&  \\
 18454& 883&4 Eri&A5IV/V      &   7740&  4.07&  3.9&   100& $-$0.20& +0.16& +0.05& +0.21& +0.74& $-$0.20&   320& $-$0.60&  \\
 76543&3561&$o^{1}$ Cnc&A5III       &   8330&  4.18&  3.9&    91& +0.00& +0.14& +0.05& +0.31& +0.92& +0.20&   321& $-$0.64&  \\
 12216& 580&50 Cas&A2V         &   9553&  3.90&  2.6&    88& $-$0.18& +0.25& +0.40& +0.15& +0.42& $-$0.04&   144& $-$0.62&  \\
 28355&1414&79 Tau&A7V         &   7809&  3.98&  4.0&    87& $-$0.06& +0.10& $-$0.35& +0.15& +0.36& $-$0.46&   287& $-$0.65& H \\
222345&8968&$\omega^{1}$ Aqr&A7IV        &   7487&  3.88&  3.8&    86& +0.01& $-$0.03& +0.03& $-$0.07& $-$0.31& $-$0.37&   310& $-$0.63&  \\
 74198&3449&$\gamma$ Cnc&A1IV        &   9381&  4.11&  2.8&    85& $-$0.32& $-$0.06& $-$0.14& +0.26& +0.64& +0.13&   186& $-$0.76&  \\
 27934&1387&$\kappa^{1}$ Tau&A7IV-V      &   8159&  3.84&  4.0&    83& +0.07& +0.12& $-$0.05& +0.00& $-$0.43& $-$0.49&   254& $-$0.64& H \\
 25490&1251&$\nu$ Tau&A1V         &   9077&  3.93&  3.2&    82& $-$0.48& $\cdots$& $-$0.44& +0.01& +0.43& $-$0.18&   179& $-$0.62&  \\
 29388&1473&90 Tau&A6V         &   8194&  3.88&  4.0&    82& +0.03& +0.08& $-$0.04& $-$0.03& $-$0.39& $-$0.47&   253& $-$0.63& H \\
 79469&3665&$\theta$ Hya&B9.5V       &  10510&  4.20&  1.4&    82& $-$0.26& $\cdots$& $\cdots$& $-$0.01& $\cdots$& $-$0.10&    71& $-$0.36&  \\
 28226&1403&         &Am          &   7361&  4.01&  3.6&    81& $-$0.07& +0.08& $-$0.38& +0.25& +0.46& $-$0.24&   322& $-$0.57& H \\
207098&8322&$\delta$ Cap&A5mF2 (IV)  &   7312&  4.06&  3.6&    81& $-$0.41& $-$0.15& $-$0.45& +0.10& +0.56& +0.06&   356& $-$0.48&  \\
216627&8709&$\delta$ Aqr&A3V         &   8587&  3.59&  3.7&    79& $-$0.27& $-$0.35& $-$0.06& $-$0.26& $-$0.20& $-$0.65&   177& $-$0.48&  \\
 33641&1689&$\mu$ Aur&A4m         &   7961&  4.21&  4.0&    79& $-$0.30& $-$0.21& $-$0.56& +0.08& +0.33& $-$0.52&   271& $-$0.61&  \\
 12111& 575&48 Cas&A3IV        &   7910&  4.08&  4.0&    76& $-$0.23& $-$0.21& $-$0.33& $-$0.26& $-$0.21& $-$0.88&   234& $-$0.59&  \\
 23281&1139&         &A5m         &   7761&  4.19&  4.0&    76& $-$0.10& $-$0.12& $-$0.44& +0.03& +0.44& $-$0.46&   291& $-$0.61&  \\
192640&7736&29 Cyg&A2V         &   8845&  3.86&  3.5&    74& $-$0.41& $-$1.21& $-$0.41& $-$1.54& $-$0.63& $-$0.86&   122& $-$0.33&  \\
 11636& 553&$\beta$ Ari&A5V...      &   8294&  4.12&  3.9&    73& $-$0.14& +0.00& $-$0.18& +0.10& +0.81& $-$0.22&   273& $-$0.64&  \\
173880&7069&111 Her&A5III       &   8567&  4.27&  3.8&    72& $-$0.05& $-$0.01& $-$0.10& +0.20& +0.95& $-$0.04&   264& $-$0.64&  \\
  5448& 269&$\mu$ And&A5V         &   8147&  3.82&  4.0&    72& $-$0.09& $-$0.14& $-$0.14& $-$0.14& $-$0.24& $-$0.48&   255& $-$0.64&  \\
 17093& 812&38 Ari&A7III-IV    &   7541&  3.95&  3.8&    69& $-$0.24& $-$0.27& $-$0.20& $-$0.25& $-$0.14& $-$0.64&   281& $-$0.65&  \\
 28319&1412&$\theta^{2}$ Tau&A7III       &   7789&  3.68&  4.0&    68& $-$0.01& +0.06& $-$0.15& $-$0.11& $-$0.43& $-$0.60&   272& $-$0.67& H \\
 95382&4294&59 Leo&A5III       &   8017&  3.95&  4.0&    68& $-$0.02& $-$0.14& $-$0.08& $-$0.10& $-$0.06& $-$0.35&   285& $-$0.65&  \\
140436&5849&$\gamma$ CrB&A1Vs        &   9274&  3.89&  3.0&    68& $-$0.33& +0.65& +0.04& $-$0.21& +0.43& $-$0.66&    99& $-$0.31&  \\
 20320& 984&$\zeta$ Eri&A5m         &   7505&  3.91&  3.8&    67& $-$0.50& $-$0.34& $-$0.76& $-$0.13& +0.22& $-$0.38&   309& $-$0.63&  \\
 13161& 622&$\beta$ Tri&A5III       &   7957&  3.68&  4.0&    65& $-$0.13& $-$0.87& $-$0.44& $-$0.40& $-$0.36& $-$0.74&   244& $-$0.63&  \\
 27045&1329&$\omega^{2}$ Tau&A3m         &   7552&  4.26&  3.8&    62& $-$0.04& +0.06& $-$0.45& +0.32& +0.80& $-$0.26&   319& $-$0.56&  \\
200499&8060&$\eta^{2}$ Cap&A5V         &   8081&  3.95&  4.0&    61& $-$0.10& $-$0.11& $-$0.10& $-$0.21& $-$0.18& $-$0.60&   250& $-$0.62&  \\
 29499&1480&         &A5m         &   7638&  4.08&  3.9&    61& $-$0.07& +0.22& +0.16& +0.32& +0.53& $-$0.22&   322& $-$0.59& H \\
116656&5054&$\zeta$ UMa&A2V         &   9317&  4.10&  2.9&    59& $-$0.48& $-$0.19& $-$0.09& +0.27& +0.70& $-$0.02&   174& $-$0.67&  \\
198639&7984&56 Cyg&A4me...     &   7921&  4.09&  4.0&    59& $-$0.19& $-$0.02& $-$0.22& +0.03& +0.27& $-$0.28&   300& $-$0.63&  \\
130841&5531&$\alpha^{2}$ Lib&A3IV        &   8079&  3.96&  4.0&    58& $-$0.64& $-$1.18& $-$1.12& $-$0.37& $-$0.16& $-$1.07&   195& $-$0.51&  \\
 30121&1511& 4    Cam&A3m         &   7700&  3.98&  3.9&    57& $-$0.58& +0.14& $-$0.57& +0.25& +0.76& $-$0.17&   325& $-$0.61&  \\
 30210&1519&         &Am...       &   7927&  3.94&  4.0&    56& $-$0.66& +0.13& $-$0.62& +0.44& +1.17& +0.12&   345& $-$0.62& H \\
 29479&1478&$\sigma^{1}$ Tau&A4m         &   8406&  4.14&  3.9&    56& $-$0.15& +0.14& $-$0.38& +0.35& +1.06& $-$0.04&   282& $-$0.66& H \\
222603&8984&$\lambda$ Psc&A7V         &   7757&  3.99&  4.0&    55& $-$0.15& $-$0.16& $-$0.13& $-$0.19& $-$0.19& $-$0.56&   280& $-$0.64&  \\
212061&8518&$\gamma$ Aqr&A0V         &  10384&  3.95&  1.5&    54& $-$0.25& $\cdots$& $\cdots$& $-$0.01& $\cdots$& +0.10&    92& $-$0.51&  \\
 89021&4033&$\lambda$ UMa&A2IV        &   8861&  3.61&  3.5&    52& $-$0.25& $-$0.07& $-$0.24& +0.06& +0.52& $-$0.19&   198& $-$0.66& (+0.21) \\
195725&7850&$\theta$ Cep&A7III       &   7816&  3.74&  4.0&    49& $-$0.38& $-$0.02& $-$0.17& +0.14& +1.02& +0.13&   351& $-$0.63&  \\
 43378&2238&2 Lyn&A2Vs        &   9210&  4.09&  3.0&    46& $-$0.13& $-$0.48& $-$0.07& $-$0.15& $-$0.18& $-$0.42&   138& $-$0.44& ($-0.14$) \\
 27819&1380&$\delta^{2}$ Tau&A7V         &   8047&  3.95&  4.0&    45& $-$0.11& $-$0.02& $-$0.01& $-$0.04& $-$0.05& $-$0.37&   279& $-$0.65& H \\
 95418&4295&$\beta$ UMa&A1V         &   9489&  3.85&  2.7&    44& $-$0.36& $-$0.20& $-$0.14& +0.21& +0.86& $-$0.03&   150& $-$0.64&  \\
218396&8799&         &A5V         &   7091&  4.06&  3.3&    41& +0.04& $-$0.60& $-$0.74& $-$0.58& $-$0.93& $-$0.94&   250& $-$0.64&  \\
 84107&3861&15 Leo&A2IV        &   8665&  4.31&  3.7&    38& $-$0.22& $-$0.26& $-$0.36& +0.01& +0.70& $-$0.31&   221& $-$0.58& (+0.01) \\
204188&8210&         &A8m         &   7622&  4.21&  3.9&    36& $-$0.13& $-$0.12& $-$0.38& +0.02& +0.32& $-$0.59&   281& $-$0.62& ($-0.04$) \\
 33204&1670&         &A5m         &   7530&  4.06&  3.8&    34& $-$0.43& +0.00& $-$0.32& +0.28& +0.98& $-$0.24&   323& $-$0.58& H \\
141795&5892&$\epsilon$ Ser&A2m         &   8367&  4.24&  3.9&    32& $-$0.69& $-$0.09& $-$0.66& +0.23& +1.15& +0.04&   297& $-$0.64&  \\
173648&7056&$\zeta^{1}$ Lyr&Am          &   8004&  3.90&  4.0&    32& $-$0.47& +0.06& $-$0.38& +0.29& +1.23& $-$0.06&   317& $-$0.66& (+0.24) \\
112185&4905&$\epsilon$ UMa&A0p         &   9407&  3.61&  2.8&    32& $-$1.13& $-$0.37& $\cdots$& +0.17& +0.38& $-$0.35&   114& $-$0.41&  \\
 27628&1368&60 Tau&A3m         &   7218&  4.05&  3.5&    30& $-$0.47& +0.09& $-$0.67& +0.32& +0.76& $-$0.06&   340& $-$0.50& H \\
 28546&1428&81 Tau&Am          &   7640&  4.17&  3.9&    28& $-$0.28& +0.02& $-$0.39& +0.22& +0.85& $-$0.36&   306& $-$0.60& (+0.10) H\\
172167&7001&$\alpha$ Lyr&A0Vvar      &   9435&  3.99&  2.7&    21& $-$0.20& $-$0.76& $-$0.47& $-$0.55& $-$0.84& $-$0.42&   113& $-$0.39&  \\
 60179&2891&$\alpha$ Gem&A2Vm        &   9122&  3.88&  3.2&    19& $-$0.40& $-$0.55& $-$0.74& $-$0.03& +0.29& $-$0.43&   142& $-$0.45&  \\
 95608&4300&60 Leo&A1m         &   8972&  4.20&  3.3&    18& $-$0.64& $-$0.04& $-$0.83& +0.30& +0.97& $-$0.06&   211& $-$0.66& (+0.17) \\
 48915&2491&$alpha$ CMa&A0m...      &   9938&  4.31&  2.1&    16& $-$0.40& $-$0.15& $-$0.52& +0.42& +1.22& +0.15&   138& $-$0.67&  \\
 27749&1376&63 Tau&A1m         &   7448&  4.21&  3.7&    13& $-$0.57& +0.05& $-$1.16& +0.47& +1.12& $-$0.12&   336& $-$0.53& (+0.23) H\\
 33254&1672&16 Ori&A2m         &   7747&  4.14&  3.9&    13& $-$0.62& +0.14& $-$0.94& +0.38& +1.02& $-$0.24&   316& $-$0.60& (+0.23) H\\
 72037&3354& 2 UMa&A2m         &   7918&  4.16&  4.0&    12& $-$0.81& +0.02& $-$1.03& +0.21& +0.71& $-$0.42&   285& $-$0.62& (+0.16) \\
 27962&1389&$\delta^{3}$ Tau&A2IV        &   8923&  3.94&  3.4&    11& $-$0.28& +0.03& $-$0.24& +0.23& +0.73& $-$0.11&   207& $-$0.68& (+0.29) H\\
 47105&2421&$\gamma$ Gem&A0IV        &   9115&  3.49&  3.2&    11& $-$0.05& $-$0.09& $-$0.04& $-$0.06& +0.20& $-$0.37&   140& $-$0.47& (+0.09) \\
 40932&2124&$\mu$ Ori&Am...       &   8005&  3.93&  4.0&    11& $-$0.54& $-$0.43& $-$0.76& $-$0.12& +0.23& $-$0.81&   234& $-$0.60& ($-0.17$) H\\
\end{longtable}

\end{document}